\documentclass[superscriptaddress, nofootinbib]{revtex4-1}[11pt]
\pdfoutput=1

\usepackage{graphicx}
\graphicspath{{figs/}}
\usepackage {amsmath, amssymb, rotating}
\usepackage{hyperref}
\usepackage[usenames,dvipsnames]{color}
\usepackage{slashed}

\newcommand{\nl}{\nonumber \\}
\newcommand{\x}{\chi}
\newcommand{\zp}{{Z^{\prime}}}

\newcommand{\nsm}{ \nu_{_\text{SM}}}
\newcommand{\order}[1]{\mathcal{O}(#1)}

\newcommand{\be}{\begin{eqnarray}}
\newcommand{\ee}{\end{eqnarray}}

\makeatletter
\def\@seccntformat#1{\csname the#1\endcsname.\quad}

\def\lsim{\mathrel{\raise.3ex\hbox{$<$\kern-.75em\lower1ex\hbox{$\sim$}}}}
\def\gsim{\mathrel{\raise.3ex\hbox{$>$\kern-.75em\lower1ex\hbox{$\sim$}}}}

\begin{document}

\hspace{11.5cm} \parbox{5cm}{FERMILAB-PUB-16-318-A}~\\
\vspace{1cm}

\title{Thermal Dark Matter From A Highly Decoupled Sector}

\author{Asher Berlin}
\thanks{ORCID: http://orcid.org/0000-0002-1156-1482}
\affiliation{Department of Physics, Enrico Fermi Institute, University of Chicago, Chicago, IL}

\author{Dan Hooper}
\thanks{ORCID: http://orcid.org/0000-0001-8837-4127}
\affiliation{Center for Particle Astrophysics, Fermi National Accelerator Laboratory, Batavia, IL 60510}
\affiliation{Department of Astronomy and Astrophysics, The University of Chicago, Chicago, IL 60637}

\author{Gordan Krnjaic}
\thanks{ORCID: http://orcid.org/0000-0001-7420-9577}
\affiliation{Center for Particle Astrophysics, Fermi National Accelerator Laboratory, Batavia, IL 60510}

\begin{abstract}
\vskip 3pt \noindent

It has recently been shown that if the dark matter is in thermal equilibrium with a sector that is highly decoupled from the Standard Model, it can freeze-out with an acceptable relic abundance, even if the dark matter is as heavy as $\sim$1-100 PeV. In such scenarios, both the dark and visible sectors are populated after inflation, but with independent temperatures. The lightest particle in the dark sector will be generically long-lived, and can come to dominate the energy density of the universe. Upon decaying, these particles can significantly reheat the visible sector, diluting the abundance of dark matter and thus allowing for dark matter particles that are much heavier than conventional WIMPs. In this paper, we present a systematic and pedagogical treatment of the cosmological history in this class of models, emphasizing the simplest scenarios in which a dark matter candidate annihilates into hidden sector particles which then decay into visible matter through the vector, Higgs, or lepton portals. In each case, we find ample parameter space in which very heavy dark matter particles can provide an acceptable thermal relic abundance. We also discuss possible extensions of models featuring these dynamics.

\end{abstract}

\maketitle


\tableofcontents


\section{Introduction}
\label{sec:intro}

Over the past several decades, weakly interacting massive particles (WIMPs) have been the leading class of candidates for our universe's dark matter. This paradigm has been motivated primarily by 
the fact that a stable particle species with a weak-scale mass and interaction strength is predicted to freeze-out of thermal equilibrium in the early universe with a relic abundance that is comparable to the measured cosmological density of dark matter. As such particles are also often found within frameworks that address the electroweak hierarchy problem (including, but not limited to, weak-scale supersymmetry), this connection has become commonly known as the ``WIMP miracle''~\cite{Bertone:2016nfn}.

The WIMP paradigm has motivated an expansive experimental program, consisting of direct detection, indirect detection, and collider searches. As these efforts have progressed, however, no conclusive detections have been made, and increasingly powerful bounds have been placed on dark matter's non gravitational interactions with the Standard Model (SM). Over the traditional range of WIMP masses ($\sim$10-1000 GeV), direct detection experiments now strongly constrain the dark matter's elastic scattering cross section with nuclei~\cite{Tan:2016zwf,Akerib:2015rjg,Akerib:2016vxi,Agnese:2015nto}, and astrophysical observations by gamma-ray telescopes~\cite{Hooper:2012sr,Ackermann:2015zua} and cosmic ray detectors~\cite{Bergstrom:2013jra,Giesen:2015ufa,Cirelli:2013hv} have also begun to constrain the WIMP parameter space. Although many WIMP models remain viable, it is perhaps surprising that no definitive detection of particle dark matter has yet been made.

In light of this experimental situation, it has become increasingly interesting to consider dark matter scenarios beyond the conventional WIMP paradigm. In this paper, we focus on dark matter candidates with negligible couplings to the SM and that reside within a sector that is thermally decoupled from the visible matter in the early universe. In doing so, we build upon our previous recent work~\cite{Berlin:2016vnh} by considering a wider range of models and discussing their phenomenology in greater detail. 


Throughout this study, we assume that the visible sector, which contains the SM, is supplemented by a decoupled hidden sector, which contains the dark matter.  We further assume that both sectors are
 thermally populated during post-inflation reheating and maintain separate temperatures throughout cosmological evolution~\cite{Allahverdi:2010xz,Adshead:2016xxj}.
 Although sequestered from the SM, the hidden sector may consist of many new additional particle species with sizable mutual interaction rates. In particular, it is possible that the lightest stable hidden species, $X$, freezes out via $XX \to YY$ annihilation, where $Y$ is a lighter hidden sector species that ultimately decays into SM particles. Being stable, we take $X$ to be our dark matter candidate.

If the $Y$ is short-lived, it will never dominate the energy density of the universe, and will have little effect on cosmological evolution. 
In this regime, $X$ will freeze out with the observed dark matter abundance only if its mass and couplings are similar to those of traditional WIMPs.  Although, in principle, such a scenario can be viable for a wide range of masses, constraints from perturbative unitarity typically require $m_X \lesssim \order{100}$ TeV~\cite{Griest:1989wd} (see, however, Ref.~\cite{Harigaya:2016nlg}). This bound can be circumvented, however, if the entropy of the visible sector increases appreciably after the freeze-out of $X$~\cite{Dev:2016xcp,Fornengo:2002db,Gelmini:2006pq,Kane:2015jia,Hooper:2013nia,Patwardhan:2015kga,Randall:2015xza,Reece:2015lch,Lyth:1995ka,Davoudiasl:2015vba,Cohen:2008nb,Yamanaka:2014pva}.
For instance, a heavy and long-lived species in the hidden sector could come to dominate the energy density of the universe before decaying to SM particles, thereby diluting all relic abundances, including that of $X$. As we will see in Sec.~\ref{sec:decay}, the increase in the visible sector entropy from $Y$ decay scales as $\propto \tau_Y^{1/2}$, where $\tau_Y$ is the lifetime of the unstable species. Thus, for sufficiently large $\tau_Y$, it is possible to significantly dilute the abundance of $X$, thereby achieving an acceptable density of dark matter, even for masses well above the conventional limit from perturbative unitarity, $m_X \gg 100$ TeV.


 Long lifetimes are straightforwardly realized if the decaying particle is the lightest hidden sector state. In fact, if the hidden and visible sectors are highly decoupled, the lightest hidden sector state will automatically be long-lived, since its width relies on a coupling that is too small to sustain thermal equilibrium between the two sectors. We emphasize that this picture is relatively universal, and can be found within any model in which the dark matter freezes out through annihilations in a heavy and highly decoupled hidden sector that is populated after inflation. In contrast to scenarios in which an additional out-of-equilibrium decay is invoked solely to dilute the initial cosmological abundances of various species, dilutions of the type considered in this paper are an inevitable consequence of thermal decoupling.

\begin{figure}[t]
\begin{center}
\includegraphics[width=0.6\textwidth]{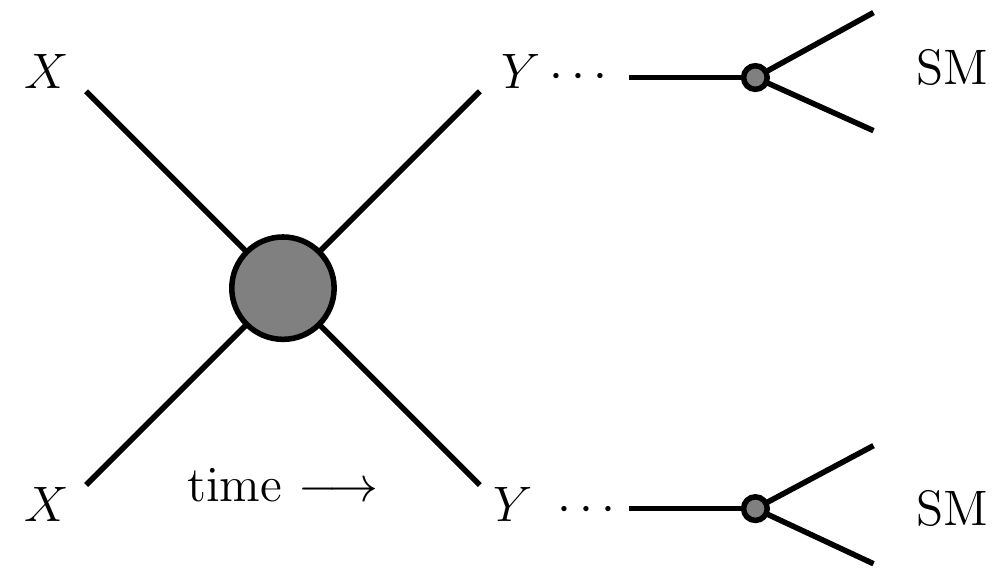}
\caption{\label{fig:schematic} A schematic diagram of the processes that we will consider in this study. Here $X$, the dark matter candidate, annihilates into pairs of metastable hidden sector $Y$ particles.  If the hidden sector is heavy and extremely decoupled from the visible sector (which contains the Standard Model), then $Y$ will be long-lived, and may eventually dominate the universe's energy density. Upon its decay into Standard Model particles, $Y$ reheats the visible universe and dilutes all particle abundances, including the relic density of $X$.}
\end{center}
\end{figure}

The remainder of this paper is structured as follows. In Sec.~\ref{sec:thermo}, we review the early universe thermodynamics of scenarios with a decoupled hidden sector. We then discuss in detail the processes of thermal freeze-out and out-of-equilibrium decay in Secs.~\ref{sec:fo} and~\ref{sec:decay}, respectively. 
In Sec.~\ref{sec:Neff}, we discuss possible contributions to the effective number of neutrino species within this class of scenarios. In Sec.~\ref{sec:models}, we describe three concrete realizations of dark matter in a decoupled hidden sector, in which the hidden and visible sectors interact through the vector portal, Higgs portal, or lepton portal. Finally, we briefly summarize our results and conclusions in Sec.~\ref{sec:conclusion}.

\section{Hidden Sector Thermodynamics}
\label{sec:thermo}

In this section, we review the thermodynamic evolution of a generic hidden sector, whose constituents interact very feebly with the visible sector. In the decoupled limit, 
these sectors influence each other's evolution only indirectly by either modifying the cosmic expansion rate, or by injecting energy through any decays of hidden sector particles into the visible sector.
We begin by considering two particle species within the hidden sector: the lightest hidden sector particle, $Y$, and the lightest stable hidden sector particle, $X$. The stable species will annihilate through processes such as $XX \to YY$ until its abundance freezes out of equilibrium, in analogy with conventional WIMP freeze-out. Since $Y$ is the lightest particle in the hidden sector, $Y$ can only decay to the SM, either directly or through a multi-step cascade, e.g., $Y \to \cdots \to \text{SM}$; this setup is depicted schematically in  Fig.~\ref{fig:schematic}. For simplicity, we will assume for the moment that the interactions between these two sectors are too feeble to reach equilibrium. Such feeble interactions could arise, e.g., through mass-mixing, loop-induced effects, or suppressed tree-level interactions, and may be sufficiently small such that $Y$ will be relatively long-lived, with a lifetime as long as $\tau_{\, Y} \sim \order{1}$ second. 

If kinetically decoupled, the hidden and visible sectors will each be described by distinct thermal distributions whose respective temperatures evolve differently over time. It is useful to define the ratio of the hidden and visible sector temperatures, $\xi \equiv T_h/T$. Here and throughout this paper, quantities pertaining to hidden sector dynamics are labelled with a subscript or superscript ``$h$", while those without such a label denote visible sector quantities. 

\begin{figure}[t]
\begin{center}
\includegraphics[width=0.7\textwidth]{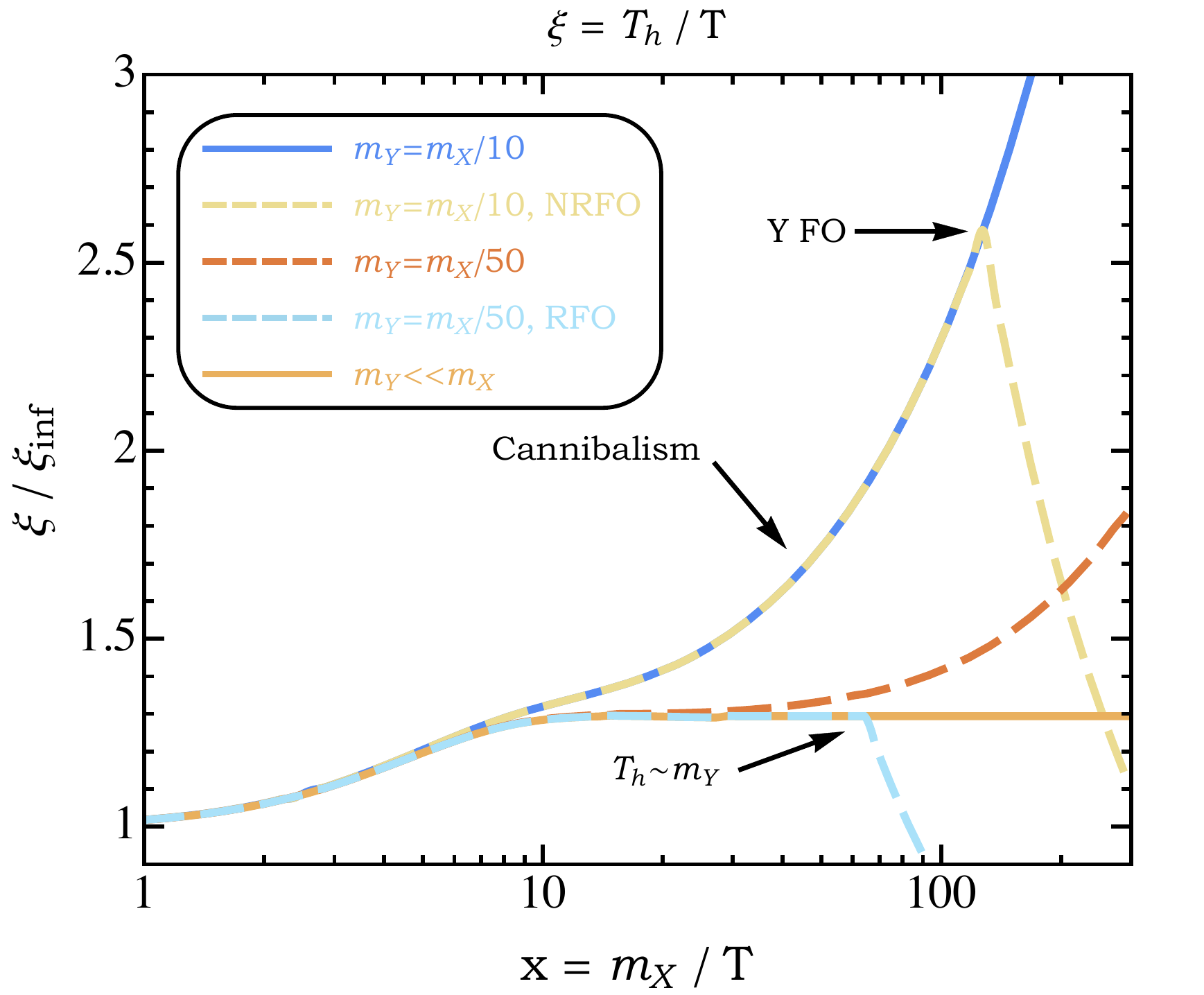}
\caption{\label{fig:xi} Temperature dependence of $\xi\equiv T_h/T$, for the case that $X$ is a Dirac fermion and $Y$ is a massive neutral vector boson. RFO (NRFO) denotes that $Y$ freezes out while (non-)relativistic. Otherwise, $Y$ is assumed to be in chemical equilibrium. Cannabilism occurs indefinitely if $Y$ remains in chemical equilibrium once $T_h \lesssim m_Y$, as seen by the sharp rise in $\xi$ for the blue and dashed-red lines, corresponding to $m_Y = m_X / 10$ and $m_Y = m_X / 50$, respectively. Similarly, for the yellow-dashed line, we once again take $m_Y = m_X / 10$, but assume that once $T_h \lesssim m_Y$, $Y$ only remains in chemical equilibrium up until it freezes out at $T_h \sim m_Y / 5$, at which point $\xi \sim 1 / R$. Also, as illustrated by the dashed light-blue line, we fix $m_Y = m_X / 50$ and assume that $Y$ freezes out while still relativistic. In this case, $\xi$ is truncated by Eq.~(\ref{eq:xi2}), up until $T_h \lesssim m_Y$, at which point $\xi \sim 1 / R$. Finally, we show the limiting case of $m_Y \ll m_X$ as depicted by the solid orange line.}
\end{center}
\end{figure}

For our initial conditions, we take $\xi = \xi_\text{inf}\, $, where the subscript denotes the value immediately following post-inflation reheating. At early times, significantly before the decay of $Y$, entropy is approximately conserved independently in both sectors. Hence, the evolution of $\xi$ can be tracked using the forms for the entropy densities, $s = (2 \pi^2 / 45) \, g_*(T) \, T^3$ and $s_h = (2 \pi^2 / 45) \, g_*^h(T_h) \, T_h^3\, $, where $g_*$ and $g_*^h$ correspond to the effective relativistic degrees-of-freedom in equilibrium with the visible and hidden sectors, respectively. Conservation of entropy implies that $s_h / s =  s_h / s |_\text{inf}\, $, from which it follows that $\xi$ evolves as
\be
\label{eq:xi1}
\xi = \left( \frac{g_*(T)}{g_{*\, \text{inf}}} \right)^{1/3} \left( \frac{g^h_{* \, \text{inf}}}{g^h_*(T_h)} \right)^{1/3} ~ \xi_\text{inf}
~.
\ee
For the most part, we will be interested in $T \gg \order{100} \text{ GeV}$, for which $g_* \simeq g_{* \, \text{inf}} \approx 106.75$. For the case of $m_Y \ll T_h \ll m_X$, $g_{* \, \text{inf}}^h = c_Y g_Y + c_X g_X$ and $g_*^h = c_Y g_Y$, where $c_{X,Y} = 1 \, (7/8)$ for bosonic (fermionic) $X$, $Y$, and $g_{X,Y}$ are the number of internal degrees-of-freedom of $X$, $Y$, respectively. Under these assumptions, Eq.~(\ref{eq:xi1}) reduces to
\be
\label{eq:xi2}
\xi = \left( 1 + \frac{c_X \, g_X}{c_Y \, g_Y} \right)^{1/3} ~ \xi_\text{inf}
~.
\ee
This behavior is exhibited in the solid orange line of Fig.~\ref{fig:xi}, corresponding to the case of $m_Y \ll m_X$, for which $\xi/\xi_{\rm inf}$ is nearly constant when $T_h \ll m_X$.

The ratio $\xi/\xi_{\rm inf}$ can evolve quite differently, however, if the conditions described above are not met; for example, if we relax the assumption that $T_h \gg  m_Y$. In this case, if $Y$ is in equilibrium for temperatures $T_h \ll m_Y$, then its entropy density is given by 
\begin{eqnarray}
s_h = \frac{m_Y n_Y}{T_h} = g_Y \bigg(\frac{m_Y^{5} T_h}{ 8\pi^{3}}\bigg)^{1/2}  \, e^{-m_{Y}/T_h}~,
\end{eqnarray}
causing the hidden sector to enter into a state of ``\emph{cannibalism}," (see, e.g., Refs.~\cite{Carlson:1992fn,Pappadopulo:2016pkp,Kuflik:2015isi,Farina:2016llk}). In this case, conservation of hidden sector entropy, $s_h R^3$, implies that
\be
R^3 ~ T_h^{1/2} ~ e^{-m_Y / T_h} = \text{constant}~,
\ee
where $R$ is the scale factor. In the limit that $T_h \ll m_Y$, the variation of the exponential dominates, and the above expression can be approximated as $e^{m_Y / T_h} \propto R^3$, or $T_h \propto m_Y / \ln{R}$. As a result, $\xi$ increases rapidly as a function of the scale factor, such that
\be
\xi \propto \frac{R}{\ln{R}}
~.
\ee
This behavior is exhibited by the solid blue, dashed red, and dashed yellow lines in Fig.~\ref{fig:xi}, each of which depict periods of cannibalism in the hidden sector.


Alternatively, if $Y$ freezes out of chemical equilibrium while still relativistic, the value of $\xi$ will be held to that described in Eq.~(\ref{eq:xi2}) until $Y$ becomes non-relativistic, at which point $\xi \propto 1 / R$. This can be seen from the phase-space density of $Y$. Once a relativistic species has frozen out in the hidden sector, its comoving number density is conserved and, as a result, the phase space density, $f$ (or equivalently $(E-\mu)/ T_h$), is held constant,
\be
f \sim e^{-(E-\mu)/T_h} \sim dn / d^3 p \sim R^{-3} / R^{-3} = \text{constant}.
\ee
Imagine that $Y$ freezes out at $T^i$, $T_h^i \gg m_Y$, and $E_Y^i \approx p_Y^i \gg m_Y$, and consider later times before $Y$ becomes non-relativistic. Assuming that $E_{Y} \gg \mu_{Y}$ and using the fact that $E_{Y}/T_h$ is fixed, the temperature of $Y$ evolves as
\be
T_h = T_h^i ~ \frac{E_Y}{E_Y^i} \approx T_h^i ~ \frac{p_Y}{p_Y^i} \approx T_h^i ~ \frac{R^i}{R} \approx T_h^i ~ \frac{T}{T^i}
~,
\ee
and, hence, $\xi = \xi^i$. Alternatively, imagine that while $Y$ is non-relativistic, its comoving number density becomes or is already fixed. In this case, its kinetic energy scales as $E_{Y, \text{kin}} \propto 1 / R^2$, and hence so does $T_h$. From this it follows that $\xi = \xi^i ~ R^i / R = \xi^i ~ T / T^i$. Furthermore, through a similar argument as above, in the non-relativistic limit,
\be
T_h = T_h^i ~ \frac{m_Y - \mu_Y}{m_Y - \mu_Y^i}
~,
\ee
which gives $\mu_Y = m_Y + (\mu^i_Y - m_Y) ~ T_h / T_h^i$.

In Fig.~\ref{fig:xi}, we illustrate the behavior of $\xi$ for a number of possible scenarios. Although we have assumed in generating this figure that $X$ is a Dirac fermion and $Y$ is a neutral vector boson, the discussion in this section is more general, and applies to $X$ and $Y$ of any spin. In evaluating $\xi$, we have numerically solved the equation $s_h (\xi \, T) / s_h (\xi_\text{inf} \, T_\text{inf}) = s (T) / s (T_\text{inf})$, along with $s = (\rho + P)/T$ and $s_h = (\rho_h + P_h)/T_h$, and the general forms for energy density and pressure of a species, $i$~\cite{Kolb:1990vq}
\begin{eqnarray}
\rho_i = \frac{g_i}{2\pi^2} \int^{\infty}_{m_i} \frac{(E^2-m^2_{i})^{1/2}}{\exp[(E-\mu_{i})/T_h] \pm 1} E^2 dE ~~,~~
P_i = \frac{g_{i}}{6\pi^2} \int^{\infty}_{m_i} \frac{(E^2-m^2_{i})^{3/2}}{\exp[(E-\mu_{i})/T_h] \pm 1} dE, 
\end{eqnarray}
where $\mu_{i}$ denotes the chemical potential and the $\pm1$ in the demoninators is positive in the case of fermions and negative for bosons.

\section{Hidden Sector Freeze-Out}
\label{sec:fo}

Chemical equilibrium in the hidden sector is governed by processes such as~ $\overbrace{Y ~Y \cdots}^{n} \leftrightarrow \overbrace{Y ~ Y \cdots}^{n-1}\, $, $XX \leftrightarrow YY$, and $X Y Y \leftrightarrow X Y$. When the rate of these reactions is overtaken by Hubble expansion, the corresponding comoving number densities become fixed (until the time at which $Y$ begins to decay). In this section, we review this process of chemical freeze-out for the case of a hidden sector that is thermally decoupled from the SM~\cite{Feng:2008mu,Cheung:2010gj,Sigurdson:2009uz}. 

The coupled system of Boltzmann equations for the number densities of $X$ and $Y$ is given by
\begin{align}
\label{eq:boltz1}
\dot{n}_X + 3 H n_X &= - \langle \sigma v \rangle_X \, \big( n_X^2 - \frac{n_Y^2}{n_Y^{\text{eq}\, 2}} \, n_X^{\text{eq} \, 2}\big) + \cdots
\nl
\dot{n}_Y + 3 H n_Y &= + \langle \sigma v \rangle_X \, \big( n_X^2 - \frac{n_Y^2}{n_Y^{\text{eq}\, 2}} \, n_X^{\text{eq} \, 2}\big) - \Gamma_Y \, n_Y - \langle \sigma v^2 \rangle_X \, \big( n_X n_Y^2 - n_Y^\text{eq} \, n_X n_Y \big)  
\nl
& ~~ - \langle \sigma v^2 \rangle_Y \, \big( n_Y^3 - n_Y^\text{eq} \, n_Y^2 \big) - \langle \sigma v^3 \rangle_Y \, \big( n_Y^4 - n_Y^{\text{eq} \, 2}  \, n_Y^2 \big) + \cdots
~,
\end{align}
where $n_{X,Y}^\text{eq}$ denotes an equilibrium number density, $\langle \sigma v \rangle_X$ is the thermally averaged cross section for $XX \to YY$, and $\Gamma_Y$ is the decay rate for $Y$ into SM particles. The quantities $\langle \sigma v^2 \rangle_X$, $\langle \sigma v^2 \rangle_Y$, and $\langle \sigma v^3 \rangle_Y$ are the thermally averaged ``\emph{cross sections}" for $XYY \to XY$, $YYY \to YY$, and $YYYY \to YY$, respectively. For brevity, we have not included symmetry factors; for example, if $X$ is not self-conjugate, then $\langle \sigma v \rangle_X$ should be replaced with $\langle \sigma v \rangle_X\, /\, 2$. The ellipses denote higher order processes that are sub-dominant. 

The Boltzmann equations in Eq.~(\ref{eq:boltz1}) are greatly simplified in the case of entropy conservation. This is valid at times significantly before the decay of $Y$, or in cases in which $Y$ never dominates the energy density. In particular, we will recast the above equations in terms of the yield or comoving number densities, $Y_{X,Y} \equiv n_{X,Y} / s$ (not to be confused with the species $Y$). Taking the time-derivative of $Y_{X,Y}$ gives
\be
\label{eq:sdotY1}
s \dot{Y}_{X,Y} = \dot{n}_{X,Y} - n_{X,Y} \, \frac{\dot{s}}{s}
~.
\ee
Conservation of visible sector entropy, $s R^3 = \text{constant}$, implies that
\be
\label{eq:sdotovers}
\frac{\dot{s}}{s} = - 3 H
~.
\ee
Then, substituting Eq.~(\ref{eq:sdotovers}) into Eq.~(\ref{eq:sdotY1}) gives
\be
\label{eq:sdotY2}
s \dot{Y}_{X,Y} = \dot{n}_{X,Y} + 3 H n_{X,Y}
~.
\ee
%
%
%
Invoking entropy conservation once again, $S \propto T^3 R^3 = \text{constant}$, gives $\dot{T}/T = - H$, which can be rewritten in terms of $x \equiv m_X / T$,
\be
\label{eq:xdot}
\dot{x} = H x
~.
\ee
Using the chain rule and Eq.~(\ref{eq:xdot}), we then have
\be
\label{eq:dotY}
\dot{Y}_{X,Y} = H \, x \, \frac{dY_{X,Y}}{dx}
~.
\ee
By substituting Eqs.~(\ref{eq:sdotY2})  and~(\ref{eq:dotY}) into Eq.~(\ref{eq:boltz1}), we find
\begin{align}
\label{eq:boltz3}
\frac{dY_X}{dx} &= \frac{- s \, \langle \sigma v \rangle_X}{H \, x}~ \big(Y_X^2 - \frac{Y_Y^2}{Y_Y^{\text{eq} \, 2}} ~ Y_X^{\text{eq} \, 2}  \big) + \cdots
\nl
\frac{dY_Y}{dx} &= \frac{1}{H \, x}~ \bigg[ s \, \langle \sigma v \rangle_X \big(Y_X^2 - \frac{Y_Y^2}{Y_Y^{\text{eq} \, 2}} ~ Y_X^{\text{eq} \, 2}  \big) - \Gamma_Y Y_Y - s^2 \langle \sigma v^2 \rangle_X \big( Y_X Y_Y^2 - Y_Y^\text{eq} Y_X Y_Y \big) 
\nl
& \qquad \qquad - s^2 \langle \sigma v^2 \rangle_Y \big( Y_Y^3 - Y_Y^\text{eq} Y_Y^2 \big) - s^3 \langle \sigma v^3 \rangle_Y \big( Y_Y^4 - Y_Y^{\text{eq} \, 2} Y_Y^2 \big) + \cdots \bigg]
~.
\end{align}
The Hubble parameter, $H$, is given in terms of the visible and hidden sector energy densities
\be
H^2 = \frac{8 \pi}{3 m_\text{pl}^2} \left( \rho + \rho_h \right) = \frac{8 \pi}{3 m_\text{pl}^2} \, \frac{\pi^2}{30} \, \left( g_* \, T^4 + g_*^h \, T_h^4 \right) \equiv \frac{4 \pi^3}{45} ~ \frac{m_X^4}{m_\text{pl}^2} ~ \frac{g_*^{\text{eff}}}{x^4}
~,
\ee
where $m_\text{pl} = 1.22 \times 10^{19}$ GeV, and we have defined $g_*^{\text{eff}} \equiv g_* + g_*^h ~ \xi^4$. 

The final abundances of $X$ and $Y$ can be found by numerically solving either Eq.~(\ref{eq:boltz1}) or~(\ref{eq:boltz3}). However, it is often the case that processes responsible for depleting the number density of $Y$ at temperatures $T_h \lesssim m_Y$ are suppressed relative to those governing the freeze-out of $X$. If there also exists the hierarchy, $m_X \gg m_Y \gg \order{100} \text{ GeV}$, it is sensible to assume that $Y$ freezes out when it is relativistic at temperatures significantly above the weak scale. Approximating $n_Y$ with the relativistic expression $n_Y \approx c_Y^\prime \zeta (3) g_Y T_h^3 / \pi^3$, where $c_Y^\prime = 1 \, (3/4)$ for bosonic (fermionic) $Y$, and $g_* = 106.75$, we have $Y_Y = Y_Y^\text{eq} \approx 0.0026 \, c_Y^\prime \, \left( g_Y + g_X \, c_X / c_Y \right) \xi_\text{inf}^3$, where we have also used Eq.~(\ref{eq:xi2}). Changing variables once again to $\Delta = Y_X - Y_X^\text{eq}$, the first line of Eq.~(\ref{eq:boltz3}) can be rewritten as
\be
\label{eq:delta}
\frac{d\Delta}{dx} = - \frac{d Y_X^\text{eq}}{dx}- f(x) \, \Delta \, \big[ \Delta + 2 \, Y_X^\text{eq} \big]
~,
\ee
where we have defined
\be
\label{eq:FofX}
f(x) \equiv \frac{s \, \langle \sigma v \rangle_X}{H \, x} = \sqrt{\frac{\pi}{45}} ~ \frac{g_*}{\sqrt{g_*^\text{eff}}} ~ m_X \, m_\text{pl} ~ \frac{a+6 \, \xi \, b / x}{x^2}
~,
\ee
and where $\sigma v_X \equiv a + b v^2$ is the cross section for $XX \to YY$ prior to thermal averaging.

It will suffice to solve Eq.~(\ref{eq:delta}) semi-analytically. To do so, first, consider its form before $X$ departs from chemical equilibrium. At this point, $Y_X$ tracks $Y_X^\text{eq}$ very closely and hence $d \Delta / d x$ is negligible, giving
\be
\label{eq:earlytimes}
\Delta = - \, \frac{dY_X^\text{eq}}{dx} ~ \frac{1}{f(x) \big[\Delta + 2 Y_X^\text{eq} \big]}
~.
\ee
Freeze-out occurs when $Y_X$ no longer tracks $Y_X^\text{eq}$, or in other words, when $\Delta$ is comparable to $Y_X^\text{eq}$. Specifically, freeze-out is defined by $\Delta = c \, Y_X^\text{eq}$,
where $c$ is some order one number chosen to match numerical solutions. We will take $c \approx 0.4$ for $s$-wave annihilation~\cite{Kolb:1990vq}. Assuming that $X$ freezes out when non-relativistic at $x = x_f$, Eq.~(\ref{eq:earlytimes}), along with $n_X^\text{eq} \approx g_X(m_X^2 / 2 \pi x_f)^{3/2} e^{-x_f}$ and $\Delta = c \, Y_X^\text{eq}$, then imply that
\be
x_f = \xi \, \ln{\left( \frac{c(c+2)}{4 \pi^3} \, \sqrt{\frac{45}{2}} \, \frac{g_X}{\sqrt{g_*^\text{eff}}} \, m_X \, m_\text{pl} \, \frac{\xi^{5/2} (a+6 \xi b / x_f)}{\sqrt{x_f} (1- 3 \xi / 2 x_f)}  \right)}
~,
\ee
where $g_*^\text{eff}$ and $\xi$ are evaluated at freeze-out. In practice, the above equation may be solved numerically for $x_f$.

Now, consider the form of Eq.~(\ref{eq:delta}) after $X$ departs from chemical equilibrium. At this point, $Y_X^\text{eq}$ is negligible due to Boltzmann suppression, and hence $\frac{d \Delta}{dx} ~ \Delta^{-2} = - f(x)$. Integrating this from $x= x_f$ to $x=\infty$, and using the fact that $ \Delta (x = \infty) \ll \Delta (x = x_f)$, we find
\be
\label{eq:Ysimple}
Y_X (x=\infty)^{-1} = \int_{x_f}^\infty \, dx \, f(x) \approx \sqrt{\frac{\pi}{45}} ~ \frac{g_*}{\sqrt{g_*^\text{eff}}} ~ m_X \, m_\text{pl} ~ \frac{a+3 \, \xi \, b / x_f}{x_f}
~,
\ee
where $g_*$ and $g_*^\text{eff}$ are evaluated at freeze-out. Note that in Eq.~(\ref{eq:Ysimple}), we have ignored variation of $g_*^\text{eff}$ from $x=x_f$ to $x = \infty$. For $\xi_\text{inf} \gg 1$, it is possible that $g_*^\text{eff}$ varies significantly over this domain, in which case we will instead use the more general form
\be
Y_X (x=\infty)^{-1} \approx \sqrt{\frac{\pi}{45}} ~ g_* ~ m_X \, m_\text{pl} ~ \int_{x_f}^\infty \, dx ~ \frac{a+ 6 \xi b /x}{x^2 \, \sqrt{g_*^\text{eff}}}
~.
\ee

The relic abundance today is evaluated as $\Omega_X  = m_X s_0 Y_X (x=\infty) / \rho_c$, where $s_0 = 2891.2$ cm$^{-3}$ is the visible sector entropy density today and $\rho_c = 1.05375 \times 10^{-5} \, h^2$ GeV cm$^{-3}$ is the critical energy density~\cite{Agashe:2014kda}. When Eq.~(\ref{eq:Ysimple}) applies, this leads to
\be
\label{eq:relicab}
\Omega_X h^2 = 8.5 \times 10^{-11} ~ \frac{x_f \sqrt{g_\star^\text{eff}}}{g_*} ~ \left( \frac{a+3 \xi b / x_f}{\text{GeV}^{-2}} \right)^{-1}
~.
\ee
This will constitute the final abundance of $X$, provided that no entropy is transferred into the visible sector. If instead the SM entropy increases by a factor $S_f / S_i$, $\Omega_X h^2$ is effectively reduced by the same factor. This is simple to see from the following argument. Imagine that the visible sector has an initial entropy of $S_i$, which is later raised to $S_f$ through some unspecified process. Before this entropy increase, $X$ has an energy density $\rho_X^i = m_X s_i Y_X$, where $s_i = S_i / R_i^3$. Expansion of the universe dilutes the energy density such that
\begin{align}
\rho_X^f &= \rho_X^i \, \frac{R_i^3}{R^3} = m_X \, Y_X \, \frac{s_i \, R_i^3}{R^3} = m_X \, Y_X \, \frac{S_i}{R^3} = m_X \, Y_X \, \frac{S_i}{R^3} \, \frac{S_f}{S_f} = \frac{m_X \, s_f \, Y_X}{S_f / S_i}
~,
\end{align}
where $s_f = S_f / R^3$. Therefore, the dark matter energy density today is $\rho_X = m_X \, s_0 \, Y_X \, / (S_f / S_i)$. Hence, $\Omega_X h^2$, as written in Eq.~(\ref{eq:relicab}), is diluted by the factor $S_f / S_i$. As we will show in the next section, the radiation coming from the late-time out-of-equilibrium decay of $Y$ naturally generates such an increase in entropy.

\section{Out-of-Equilibrium Decay}
\label{sec:decay}

In the previous section, we described the thermal freeze-out of a dark matter candidate, $X$, which resides in a sector that is highly decoupled from the SM. We now turn our attention to the lightest particle species in the hidden sector, $Y$, which is assumed to be unstable and will eventually decay into SM particles. Due to the highly decoupled nature of the hidden sector, however, we expect such decays to be highly suppressed, leading $Y$ to be long-lived. Furthermore, upon becoming non-relativistic, the energy density of $Y$ scales as $\rho_Y \propto R^{-3}$, while the visible bath instead evolves as $\rho_\text{SM} \propto R^{-4}$. As a result, $\rho_Y / \rho_\text{SM}$ scales linearly with $R$, thus making it possible for the $Y$ population to come to dominate the energy density of the early universe, and significantly reheating the SM bath upon its eventual decay. In this section, we investigate the consequences arising from this out-of-equilibrium decay, closely following the approach described in Ref.~\cite{Kolb:1990vq}.

Using the sudden-decay approximation, it is simple to work out an estimate for the reheating of the visible sector. Imagine that $Y$, which is non-relativistic, comes to dominate the energy density of the universe up until time $t=\tau_Y$, at which point it decays into SM particles which quickly thermalize with the visible bath. Using conservation of energy, the energy density of the universe immediately prior to the decay, $\rho_Y$, should equal the energy density in radiation immediately after the decay. We will denote these two snapshots in time as $t = \tau_Y - \epsilon_t$ and $t = \tau_Y + \epsilon_t$, respectively, where $\epsilon_t$ is some small positive time-scale relative to $\tau_Y$. We will also use notation such that the label ``$i$" corresponds to $t = \tau_Y-\epsilon_t$, while ``$f$" corresponds to $t = \tau_Y+\epsilon_t$. Immediately prior to decay, the Friedmann equation gives
\be
\label{eq:Hdecay}
H^2(t = \tau_Y - \epsilon_t) = \frac{4}{9 \, \tau_Y^2} = \frac{8 \pi}{3\, m_\text{pl}^2} \, \rho_Y = \frac{8 \pi}{3\, m_\text{pl}^2} \, s_i m_Y Y_Y = \frac{16 \pi^3}{135 \, m_\text{pl}^2} \, g_* T_{i}^3 m_Y Y_Y
~,
\ee
or equivalently, 
\be
\label{eq:TRi}
T_{i}^3 = \frac{15 m_\text{pl}^2}{4 \pi^3 g_* m_Y Y_Y \tau_Y^2}
~.
\ee
Solving for $\rho_Y$ in terms of $\tau_Y$ in Eq.~(\ref{eq:Hdecay}) and enforcing energy conservation leads to
\be
\rho_Y = \frac{m_\text{pl}^2}{6 \pi \tau_Y^2} = \frac{\pi^2}{30} \, g_* \, T_{f}^4 
~,
\ee
or equivalently for the reheat temperature,
\be
\label{eq:TRf}
T_{f}^3 = \left( \frac{5 m_\text{pl}^2}{g_* \pi^3 \tau_Y^2} \right)^{3/4}
~.
\ee
The increase in SM entropy, in the sudden-decay approximation, is then found by taking the ratio of $T_{f}^3/ T_{i}^3$,
\be
\label{eq:suddendecay}
\frac{S_f}{S_i} = \frac{T_{f}^3}{T_{i}^3} \approx 2.1 ~ g_*^{1/4} ~ \frac{m_Y Y_Y \tau_Y^{1/2}}{m_\text{pl}^{1/2}}
~.
\ee

We will now derive the change in entropy more systematically, no longer relying on the sudden-decay approximation. From the definition of $\tau_Y$, $N_Y \propto e^{-t / \tau_Y}$, we obtain the differential equation, 
\be
\frac{d(R^3 n_Y)}{dt} = - \frac{1}{\tau_Y} \, R^3 n_Y
~,
\ee
which when expanded and divided by $R^3$ gives
\be
\dot{n}_Y + 3 H n_Y = - n_Y / \tau_Y
~.
\ee
Since $Y$ is assumed to be non-relativistic, $\rho_Y = m_Y n_Y$, and the above equation is equivalent to
\be
\label{eq:doe1}
\dot{\rho}_Y + 3 H \rho_Y = - \rho_Y / \tau_Y
~,
\ee
which has the general solution
\be
\label{eq:rho1}
\rho_Y (R) = \rho_Y (R_i) \left( \frac{R_i}{R} \right)^3 e^{-(t-t_i) / \tau_Y}
~.
\ee

Now, imagine that as $Y$ decays, the energy deposited is rapidly converted into relativistic thermalized particles. It follows from the second law of thermodynamics that
\be
dS = \frac{dQ}{T} = \frac{-d(R^3 \rho_Y)}{T} = \frac{-R^3}{T} \, dt \, (\dot{\rho}_Y + 3 H \rho_Y) = \frac{R^3 \rho_Y}{T} \, (dt / \tau_Y)
~,
\ee
where in the last equality we have used Eq.~(\ref{eq:doe1}). Solving $S = (2\pi^2 / 45) g_* T^3 R^3$ for $T$ and substituting into the equation above, 
\begin{align}
\label{eq:doe2}
S^{\, 1/3} \, \dot{S} = S^{\, 1/3} \,\frac{R^3}{T} \, \frac{\rho_Y}{\tau_Y} = \left( \frac{2 \pi^2}{45} g_* \right)^{1/3} \, \frac{R^4 \rho_Y}{\tau_Y} = \left( \frac{2 \pi^2}{45} g_* \right)^{1/3} \, \frac{R \, R_i^3}{\tau_Y} ~ \rho_Y (R_i)  ~ e^{-(t-t_i) / \tau_Y}
~,
\end{align}
where in the last equality we used Eq.~(\ref{eq:rho1}). A formal solution to Eq.~(\ref{eq:doe2}) is
\begin{align}
\label{eq:S}
S^{4/3} &= S_i^{4/3}  + \frac{4}{3} \, \rho_Y (R_i) R_i^4 ~~\tau_Y^{-1} \int_{t_i}^t  \, d t^\prime \, \left( \frac{2 \pi^2}{45} g_* \right)^{1/3} \, \frac{R(t^\prime)}{R_i}  \, e^{-(t^\prime-t_i) / \tau_Y}
\nl
&\equiv S_i^{4/3}  + \frac{4}{3} \, \rho_Y (R_i) R_i^4 ~ I
~.
\end{align}
To simplify Eq.~(\ref{eq:S}), we take note of two important relations involving the energy density of SM radiation, $\rho_R$, and the visible sector entropy, $s\, $:
\be
\label{eq:rhoR1}
s \, T = \frac{4}{3} \, \rho_R
~,
\ee
and
\be
\label{eq:rhoR2}
\rho_R = \frac{3}{4}  \left( \frac{45}{2 \pi^2 g_*} \right)^{1/3} S^{4/3} \, R^{-4}
~.
\ee
We then have
\begin{align}
\label{eq:rhoR3}
\rho_Y (R_i) R_i^4 S_i^{-4/3} &= m_Y R_i Y_Y S_i^{-1/3} 
\nl
&= m_Y R_i Y_Y S_i^{-1/3} ~ \times ~  \frac{4 \rho_R (R_i) / 3}{S_i T_i / R_i^3}
\nl
&= m_Y R_i Y_Y S_i^{-1/3} ~ \times ~ \frac{1}{S_i T_i / R_i^3} ~ \times ~ \left( \frac{45}{2 \pi^2 g_*(T_i)} \right)^{1/3} ( S_i^{4/3} / R_i^4 )
\nl
&= \frac{m_Y Y_Y}{T_i}  \left( \frac{45}{2 \pi^2 g_*(T_i)} \right)^{1/3}
~,
\end{align}
where we used Eq.~(\ref{eq:rhoR1}) and Eq.~(\ref{eq:rhoR2}) in the second and third lines, respectively. Taking $t_i << \tau_Y$, $t_f \gg \tau_Y$ and substituting Eq.~(\ref{eq:rhoR3}) into Eq.~(\ref{eq:S}) then implies
\be
\label{eq:S2}
\frac{S_f}{S_i} = \left[ 1 + \frac{4}{3}  \left( \frac{45}{2 \pi^2 g_*(T_i)} \right)^{1/3}  \, \frac{m_Y Y_Y}{T_i} \, I \right]^{3/4}
~,
\ee
where now $I$ is defined to be
\be
\label{eq:I_int}
I \equiv \tau_Y^{-1} \int_0^\infty dt \left( \frac{2 \pi^2}{45} g_* \right)^{1/3} \frac{R(t)}{R(0)} e^{-t/\tau_Y}
~.
\ee

In the limit that $Y$ dominates the energy density before its decay, a numerical form of $I$ is sufficient and Eq.~(\ref{eq:S2}) can be approximated as
\be
\frac{S_f}{S_i} \approx 1.83 ~ \langle g_*^{1/3} \rangle^{3/4} \, \frac{m_Y Y_Y \tau_Y^{1/2}}{m_\text{pl}^{1/2}}
~,
\ee
where the brackets indicate time-averaging over the decay~\cite{Kolb:1990vq}. Note that the difference between this numerical solution and that found using the sudden-decay approximation is at most $\order{1}$. In practice, throughout this study, we will numerically solve the system of equations, consisting of Eqs.~(\ref{eq:doe2}), (\ref{eq:S2}), and~(\ref{eq:I_int}), and the Friedmann equation,
\be
H^2 = \frac{8 \pi}{3 m_\text{pl}^2} \left( \rho_X + \rho_Y + \rho_R \right)
~,
\ee
where $\rho_Y$ is determined from Eq.~(\ref{eq:rho1}), $\rho_R = \pi^2 g_* T^4 / 30$, $\rho_X \propto R^{-3}$, and $S = (2 \pi^2 / 45) g_* T^3 R^3$. 

\section{The Effective Number of Neutrino Species}
\label{sec:Neff}

In models with a decoupled sector, there may be additional relativistic particles present during or after Big Bang Nucleosynthesis (BBN), with the potential to impact the measured expansion history of the universe. In this section, we briefly discuss this possibility within the context of the class of models under consideration here.

In generality, the effective number of neutrino species, $N_\text{eff}$, is defined in terms of the energy density of the universe, or equivalently in terms of $g_*^\text{eff}$. Allowing the neutrino temperature to be different than that of the SM plasma, we have
\be
g_*^\text{eff} = g_*^{\text{SM} - \nu} + g_*^\nu \, \xi_\nu^4 + g_*^h \, \xi_h^4 \equiv g_*^{\text{SM} - \nu} + \frac{7}{8} \times 2 \times N_\text{eff} \times ( \xi_\nu^0 )^4
~,
\ee
where $\text{SM} - \nu$ denotes the SM \emph{omitting} the three species of neutrinos, $\xi_\nu \equiv T_\nu / T$, $\xi_h \equiv T_h / T$~(we have restored the $h$ subscript for clarity), and $\xi_\nu^0$ is $T_\nu / T$ in the SM when neutrino reheating from electron-positron annihilations is neglected, i.e., $\xi_\nu^0 = (4 / 11)^{1/3} \approx 0.714$ for $T \lesssim m_e$ and $\xi_\nu^0 = 1$ for $T \gtrsim m_e$.

For $n_\nu$ flavors of neutrinos, we have
\be
\frac{7}{8} \times 2 \times N_\text{eff} \times ( \xi_\nu^0 )^4 = \frac{7}{8} \times 2 \times n_\nu \times \xi_\nu^4 ~ + ~ g_*^h \, \xi_h^4
~.
\ee
Solving for $N_\text{eff}$ yields
\be
N_\text{eff} = n_\nu \left( \frac{\xi_\nu}{\xi_\nu^0} \right)^4 + \frac{4}{7} ~ g_*^h ~ \left( \frac{\xi_h}{\xi_\nu^0} \right)^4
~.
\ee
In the SM, $g_*^h = 0$, $n_\nu = 3$, and when $T \lesssim m_e$, $\xi_\nu$ is slightly larger than $\xi_\nu^0$, so that $N_\text{eff} \approx 3.046$.

Consider the case of 3 neutrino flavors and standard cosmology ($\xi_\nu = \xi_\nu^0$) with an additional decoupled hidden sector. At early times, around BBN, for example, $T_\nu = T_\gamma$ and so the analogous calculation yields
\be
N_\text{eff} \approx 3 + \frac{4}{7} ~ g_*^h ~ \xi_h^4 ~~~ \text{(BBN)}
~.
\ee
Alternatively, after neutrino decoupling, for instance at recombination, 
\be
N_\text{eff} \approx 3.046 + \frac{4}{7} ~ \left(\frac{11}{4}\right)^{4/3} g_*^h ~ \xi_h^4 ~~~ \text{(CMB)}
~,
\ee
in agreement with Ref.~\cite{Feng:2008mu}.


Alternatively, we can also consider contributions to $N_{\rm eff}$ that arise from the decay products of the long-lived particle species, $Y$. More specifically, consider a scenario in which $Y$ has a finite branching fraction, $B_a$, to a light and decoupled state, $a$. For as long as this population of decay products remains relativistic, they will continue to contribute to $N_{\rm eff}$ (after which they will behave like matter). This will be the case so long as $T \gg T_{f} m_a / f_a m_Y$, where $T_{f}$ is the temperature of the universe following the decays of $Y$ and $f_a$ is fraction of $m_Y$ that goes into an individual $a$ (for example, for $Y \rightarrow aa$, $f_a=0.5$).

Including the contribution from these decay products, the effective number of neutrino species is given by
\begin{equation}
N_{\rm eff} \approx 3.046 + \frac{43}{7} \bigg(\frac{B_a}{1-B_a}\bigg) \bigg(\frac{g_{\star}(T_{\nu, {\rm dec}})}{g_{\star} (T_{f})}\bigg)^{1/3},
\end{equation}
where $g_{\star}(T_{\nu, {\rm dec}}) \approx 10.75$ and $T_{\nu, {\rm dec}}$ is the temperature at neutrino decoupling. Comparing this expression to constraints on $N_{\rm  eff}$ from measurements of the CMB ($N_{\rm eff} = 3.15 \pm 0.23$)~\cite{Ade:2015xua}, we conclude that $B_a \lsim 0.1 \, [g_{\star}(T_{f})/100]^{1/3}$. Next generation CMB experiments are anticipated to improve significantly upon this constraint~\cite{Dodelson:2013pln,Feng:2014uja,Wu:2014hta}.

\section{Models of Dark Matter in a Decoupled Sector}
\label{sec:models}

The scenario described above is generic and can be applied to several different classes of models. If the hidden sector is composed of SM gauge singlets, it is natural for it to be very weakly coupled to the visible bath. However, in order to facilitate the decay of the metastable state, $Y$, into SM particles, some portal between the two sectors must be introduced. At the renormalizable level, such decays can proceed through the following  three operators: $B_{\mu \nu}$, $|H|^2$, and $HL$, known as the vector, Higgs, and lepton portals, respectively. 

In this section, we will investigate models utilizing each of these portals in turn, focusing on the phenomenology outlined in Secs.~\ref{sec:thermo}-\ref{sec:decay}. Each model contains unique features and introduces complications beyond the simplest possible realization. We will proceed in order of increasing complexity. In particular, in Sec.~\ref{sec:vector} we explore the vector portal, which serves as a simple and concrete manifestation of the generic scenario described in the previous sections. In Sec.~\ref{sec:higgs}, we proceed to the Higgs portal, which necessitates a careful treatment of the freeze-out process, due to the fact that the singlet-like scalar mediator may remain in chemical equilibrium while non-relativistic. Sec.~\ref{sec:lepton} presents the lepton portal model, whose ultraviolet structure incorporates a heavy right-handed neutrino which may have potential implications for leptogenesis. 

\subsection{Vector Portal}
\label{sec:vector}

In the vector portal scenario~\cite{Pospelov:2007mp,Krolikowski:2008qa}, a new spontaneously broken $U(1)_X$ gauge symmetry is introduced, along with a corresponding massive neutral gauge boson, $Z^\prime$. As our dark matter candidate, we add to this model a complex scalar, $\phi$, which has a unit charge under $U(1)_X$ and couples to the $Z^\prime$ through the gauge coupling $g_{Z^\prime}$. $\phi$ does not acquire a vacuum expectation value (VEV) and is independent of the breaking of $U(1)_X$. Alternatively, one could also consider dark matter in the form of a Dirac fermion, as we explored previously in Ref.~\cite{Berlin:2016vnh}. If there exist particles charged under $U(1)_X \times U(1)_Y$, a small degree of kinetic mixing between the $Z^\prime$ and the SM hypercharge gauge boson can be radiatively generated. The hidden sector Lagrangian then contains the following interactions
\be
\mathcal{L} \supset - \frac{\epsilon}{2} \, B^{\mu \nu} Z_{\mu \nu}^\prime + i g_{Z^\prime} ~ Z_\mu^\prime (\phi^* \partial_\mu \phi - \phi \partial_\mu \phi^*) + g_{Z^\prime}^2 ~ Z^{\prime \mu} Z_\mu^\prime ~ |\phi|^2
~.
\ee
There may also exist direct couplings between $\phi$ and the SM Higgs through the interaction, $|\phi|^2 |H|^2$. However, since the hidden and visible sectors are thermally decoupled, this interaction must be significantly suppressed. In this section, we take the kinetic mixing parameter, $\epsilon$, to be the only relevant coupling between the two sectors.

In the limit that $m_{Z^\prime} \gg m_Z$, mixing through $\epsilon$ generates an effective interaction between the $Z^\prime$ and SM fermions,
\be
\mathcal{L} \supset - \epsilon \, g_1 \,  \sum\limits_{f} \, Y_f ~ Z_\mu^\prime ~ \bar{f} \gamma^\mu f + \order{m_Z / m_{Z^\prime}}
~,
\ee
where $g_1$ is the hypercharge gauge coupling and $Y_f$ is the hypercharge of the SM fermion, $f$~\cite{Hoenig:2014dsa}. This allows the $Z^\prime$ to decay to SM fermions with a width given by 
\be
\label{eq:gammaZp}
\Gamma_{Z^\prime} = \frac{5}{3} \, \alpha_1 \, \epsilon^2 \, m_{Z^\prime} + \order{m_Z / m_{Z^\prime}}
~.
\ee
Similarly, $\phi$ couples to the SM $Z$ through the terms
\be
- \mathcal{L} \supset \frac{i g_{Z^\prime} \epsilon s_w m_Z^2}{m_{Z^\prime}^2} ~ Z_\mu (\phi^* \partial_\mu \phi - \phi \partial_\mu \phi^*) + \frac{2 g_{Z^\prime}^2 \epsilon s_w m_Z^2}{m_{Z^\prime}^2} ~ Z^{\prime \mu} Z_\mu ~ |\phi|^2 + \order{\epsilon^2}
~,
\ee
where $s_w$ is sine of the Weinberg angle. Through $Z$ and $Z^\prime$ exchange, these interactions allow $\phi$ to scatter off protons in underground direct detection experiments, leading to a spin-independent cross section given by
\be
\sigma_p = 4 ~ g_1^2 \, c_w^4 \, \alpha_X \, \epsilon^2 ~ \frac{\mu^2}{m_{Z^\prime}^4}
~,
\ee
where $\alpha_X \equiv g_{Z^\prime}^2 / 4 \pi$, $\mu$ is the reduced mass of the proton and $\phi$, and $c_w$ is cosine of the Weinberg angle. 

Before any large increase in entropy occurs from $Z^\prime$ decays, $\phi$ freezes out through the process $\phi \bar{\phi} \to Z^\prime Z^\prime$, with an initial abundance given by Eq.~(\ref{eq:relicab}). In particular, $\frac{1}{2} \, \sigma v (\phi \bar{\phi} \to Z^\prime Z^\prime) = a + b \, v^2$, where
\begin{align}
a &= \frac{\pi  \alpha_X^2}{2 m_\phi^2} ~ \sqrt{1-r^2} ~ \bigg( 2 + \frac{r^4}{\left(2 - r^2\right)^2} \bigg) \approx  \frac{\pi \alpha_X^2}{m_\phi^2} + \mathcal{O}(r^2)~,
\nl
\nl
b &= \frac{\pi  \alpha_X^2}{48 m_\phi^2} ~ \bigg(\frac{27 r^{10}-254 r^8+900 r^6-1528 r^4+1312 r^2-448}{\left(1-r^2\right)^{1/2} \left(2 - r^2\right)^4} \bigg) \approx  -\frac{7 \pi \alpha_X^2}{12 m_\phi^2} + \mathcal{O}(r^2)
~,
\end{align}
where $v$ is the relative $\phi$ velocity, and $r \equiv m_{Z^\prime} / m_\phi \, $. 

\begin{figure}[t]
\begin{center}
\includegraphics[width=1\textwidth]{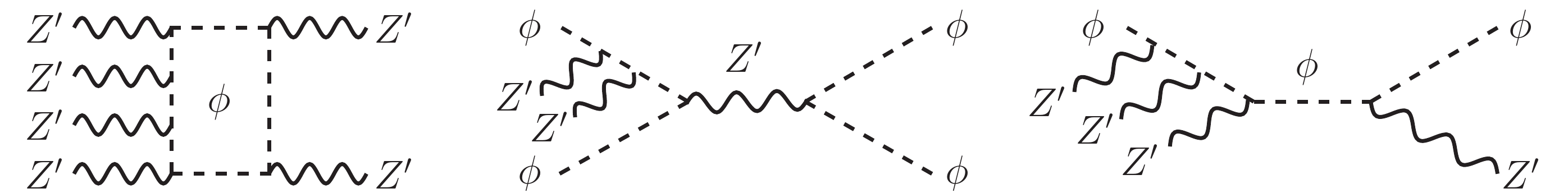}
\caption{\label{fig:Zp_Feynman} Representative Feynman diagrams for processes that could potentially maintain the chemical equilibrium of the $Z^\prime$ population for $T_h \lesssim m_{Z^\prime}$.}
\end{center}
\end{figure}

The dilution of the $\phi$ density from late-time $Z^\prime$ decays directly follows the discussion in Sec.~\ref{sec:decay}. As seen from Eq.~(\ref{eq:suddendecay}), the required inputs are $\tau_{Z^\prime}$ and $Y_{Z^\prime}$, the former of which is given by the inverse of Eq.~(\ref{eq:gammaZp}). Various processes may keep $Z^\prime$ in chemical equilibrium (with respect to the rest of the hidden sector) as the hidden sector cools. Representative diagrams that deplete the $Z^\prime$ number density are shown in Fig.~\ref{fig:Zp_Feynman}. In the discussion preceding Eq.~(\ref{eq:delta}), we noted that solving the Boltzmann equation is immensely simplified if the $Z^\prime$ departs from chemical equilibrium while it is still relativistic. Alternatively, in order for the $Z^\prime$ to remain in chemical equilibrium while non-relativistic, the rate, $\Gamma$, for a process that depletes the $Z^\prime$ number density must overcome Hubble expansion at or before the critical temperature, $T_h = m_{Z^\prime}$. Therefore, the quantity of interest is $\Gamma / H$, as evaluated at $T_h = m_{Z^\prime}$. If $\Gamma / H \ll 1$, it is safe to assume that the $Z^\prime$ population freezes out while still relativistic. 

\begin{figure}[t]
\begin{center}
\includegraphics[width=0.497\textwidth]{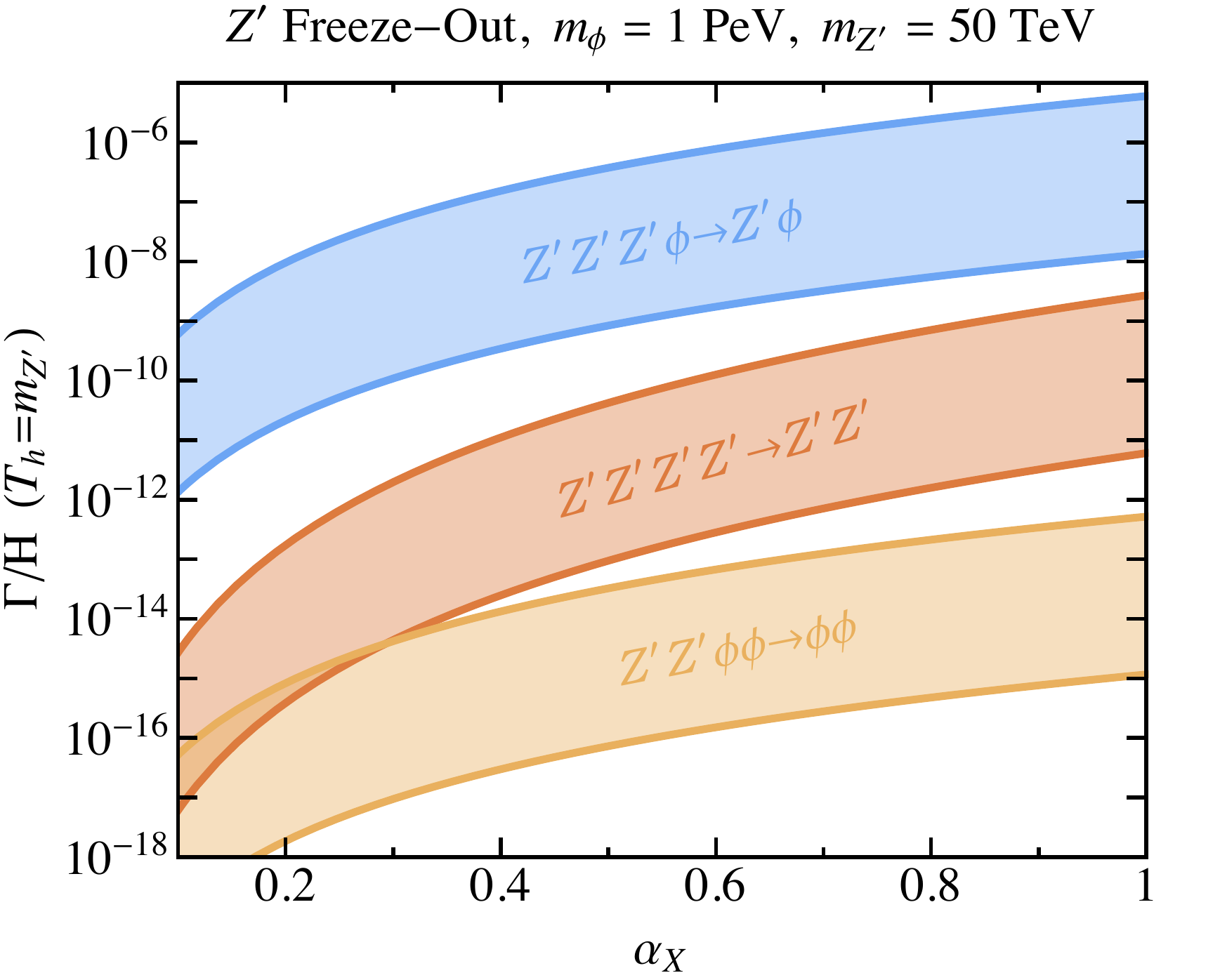}
\includegraphics[width=0.497\textwidth]{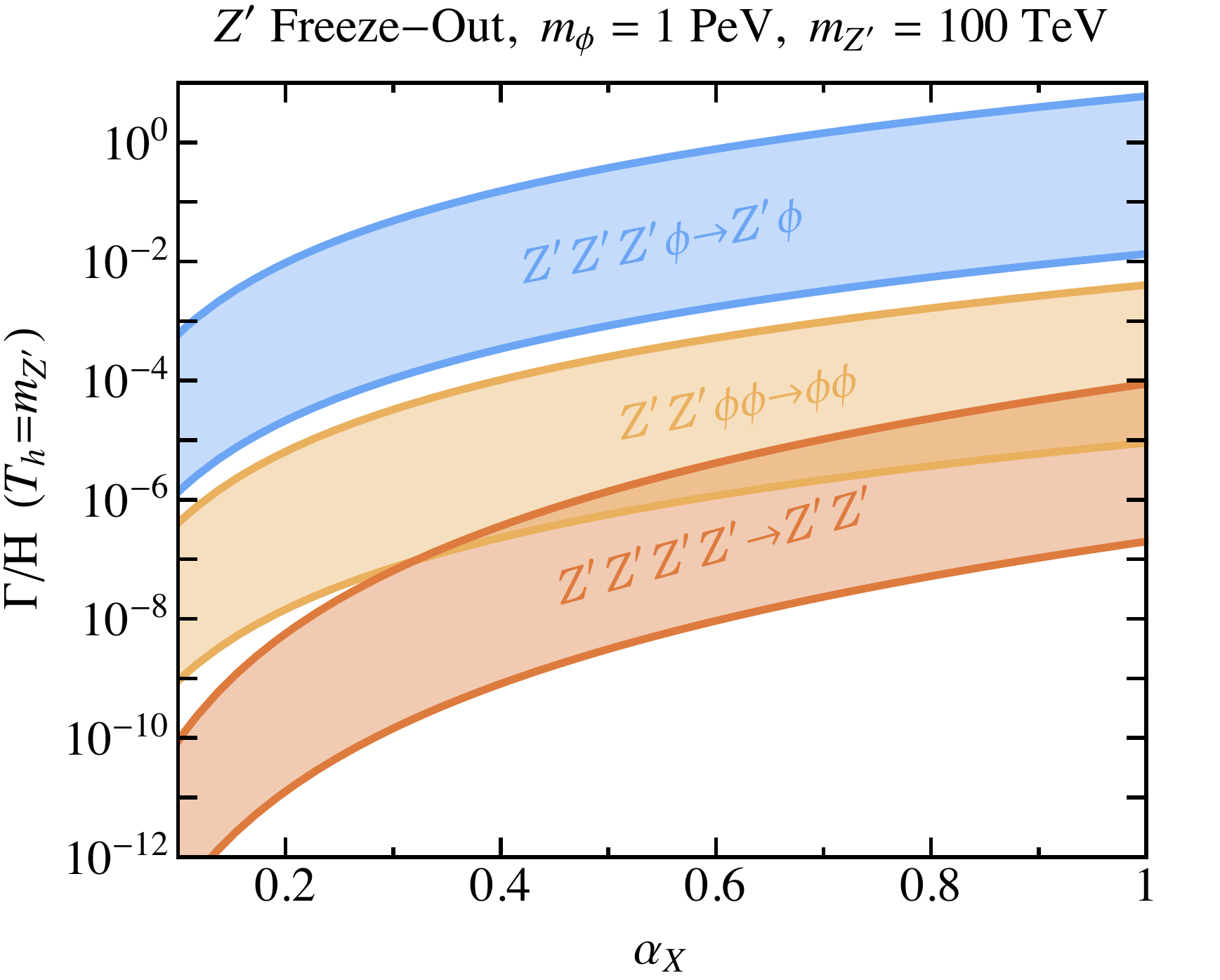}
\caption{\label{fig:Zp_FO} $\Gamma / H$ evaluated at $T_h = m_{Z^\prime}$ as a function of the coupling $\alpha_X$, for the processes $Z^\prime Z^\prime Z^\prime Z^\prime \to Z^\prime Z^\prime$ (red), $Z^\prime Z^\prime \phi \phi \to \phi \phi$ (orange), and $Z^\prime Z^\prime Z^\prime \phi \to Z^\prime \phi$ (blue), assuming that the hidden and visible sectors are thermally decoupled. We have taken $m_\phi = 1$ PeV, and $m_{Z^\prime}= 50 \, (100)$ TeV in the left (right) panels. The width of the bands corresponds to $\xi_{\text{inf}}=0.1 - 10$. Larger values of $\xi_\text{inf}$ lead to larger rates relative to that of Hubble expansion. For $m_{Z^\prime}= 50$ TeV, corresponding to the left panel above, $\Gamma / H \ll 1$ and hence the $Z^\prime$ population departs from chemical equilibrium while still relativistic. For smaller ratios of $m_\phi / m_{Z^\prime}$, corresponding to the right panel, processes that deplete the $Z^\prime$ number density allow the $Z^\prime$ to remain in chemical equilibrium while non-relativistic for $\alpha_X \gtrsim 0.5$.}
\end{center}
\end{figure}

We first consider the process $Z^\prime Z^\prime Z^\prime Z^\prime \to Z^\prime Z^\prime$ mediated by a $\phi$ loop. 
Gauge invariance and dimensional analysis suggests that the rate for this process will scale as follows:
\be
\Gamma (Z^\prime Z^\prime Z^\prime Z^\prime \to Z^\prime Z^\prime) \sim n_{Z^\prime}^3 \, \frac{\alpha_X^6 m_{Z^\prime}^8}{m_\phi^{16}}
~.
\ee
Similarly, the rates for the tree-level processes $Z^\prime Z^\prime \phi \phi \to \phi \phi$ and $Z^\prime Z^\prime Z^\prime \phi \to Z^\prime \phi$ can be written as
\be
\Gamma (Z^\prime Z^\prime \phi \phi \to \phi \phi) \sim n_{Z^\prime} n_\phi^2 \, \frac{\alpha_X^4}{m_\phi^8} ~ ,\quad \Gamma (Z^\prime Z^\prime Z^\prime \phi \to Z^\prime \phi) \sim n_{Z^\prime}^2 n_\phi \, \frac{\alpha_X^4}{m_\phi^8}
~.
\ee
In Fig.~\ref{fig:Zp_FO}, we plot the quantity $\Gamma / H $, evaluated at $T_h = m_{Z^\prime}$, as a function of $\alpha_X$ for each of these three interactions. As illustrated in this figure, for $\alpha_X \lesssim 0.5$ and $m_\phi / m_{Z^\prime} \gtrsim 10$, $\Gamma / H \lesssim 10^{-1}$, and the $Z^\prime$ population is not maintained in chemical equilibrium. For the remainder of our analysis, we will therefore assume that the $Z^\prime$ freezes out while it is relativistic. Following the discussion above Eq.~(\ref{eq:delta}), this implies that the $Z^\prime$ comoving number density is $Y_{Z^\prime} \approx 0.013 ~ \xi_\text{inf}^3 \, $. Assuming that $Z^\prime$ freezes out while relativistic allows us to focus solely on the first line of Eq.~(\ref{eq:boltz3}). Despite this simplification, the term proportional to $( Y_{Z^\prime}/ Y_{Z^\prime}^\text{eq} )^2$ deviates from unity when $T_h \lesssim m_{Z^\prime}$ and the equilibrium comoving number density becomes Boltzmann suppressed. By numerically solving the Boltzmann equation, we find that the inclusion of this effect alters our results by $\order{5 \%}$ for $m_{\phi} / m_{Z^\prime} \approx 5$ and by only $\order{1 \%}$ for $m_{\phi} / m_{Z^\prime} \approx 20$, relative to that obtained using the semi-analytic approximation.

Throughout, we have assumed that $\epsilon$ is sufficiently small such that $\phi$ and $Z^\prime$ are thermally decoupled from the SM bath. We now revisit this assumption, and consider scattering processes that could potentially equilibrate the two sectors for sufficiently large values of $\epsilon$. The dominant interactions are $Z^\prime f \to (\gamma / g) f$ and $( \gamma / g) f \to Z^\prime f$, where $f$ is some SM fermion. At leading order in $m_f / m_{Z^\prime}$, we find
\begin{align}
\hspace{-0.1cm}\sigma v (Z^\prime f \to \gamma f) &\approx \frac{\alpha_\text{em} \, Q_f^2 \, \left(g_v^2+g_a^2\right)}{6 \left(s-m_\zp^2\right)^2} \,  \Bigg[ s+ 6 m_\zp^2 -
\frac{7 m_\zp^4 }{ s} + 2 \left( s - 2 m_\zp^2 + \frac{2 m_\zp^4 }{ s} \right) \log{\frac{s (1-m_\zp^2/s )^2}{m_f^2}} \Bigg]~,
\end{align}
while for the reverse process 
\begin{align}
\sigma v (\gamma f \to Z^\prime f) &= \frac{3(s-m_\zp^2)^2}{2s^2} \sigma v (Z^\prime f \to \gamma f)~. 
%
\end{align}
Here $\sqrt{s} \approx 4 T$ is the center of mass energy\footnote{For thermal distributions of bosons and fermions, the average energy per particle is approximately $\rho/n = 2.70~T$ and $3.15~T$ respectively, so 
for fermion-boson scattering, the angle averaged $s = (p_1 + p_2)^2 \to 2 E_1 E_2 \approx (4 T)^2$.}, $Q_f$ is the electric charge of $f$, and $g_{v,a} = -\epsilon \, g_1 \, (Y_{f_R} \pm Y_{f_L}) \, $, where $Y_{f_{L/R}}$ is the hypercharge  of SM fermion $f_{L/R}$. For processes involving gluons instead of photons, one simply replaces the quantity $\alpha_\text{em} \, Q_f^2$ with $4 \, \alpha_s$. If $n_f \, \sigma v \lesssim H$ at $T = m_\phi / x_f$, then the hidden and visible sectors do not equilibrate before the freeze-out of the dark matter abundance. 

For our numerical results, we include contributions from all SM fermions, $f$, and all gauge interactions involving gluons and electroweak gauge bosons. We safely neglect contributions from pure gauge boson external states ({\it i.e.} $Z^\prime \gamma \to ff$) since, for $T \gg v$, these are highly subdominant to the total contribution from $Z^\prime f$ initiated rates and the corresponding reverse processes on account of $g_*(T) \simeq 100$.

\begin{figure}[h!]
\begin{center}  \hspace{-0.7cm} 
\includegraphics[width=3.2in]{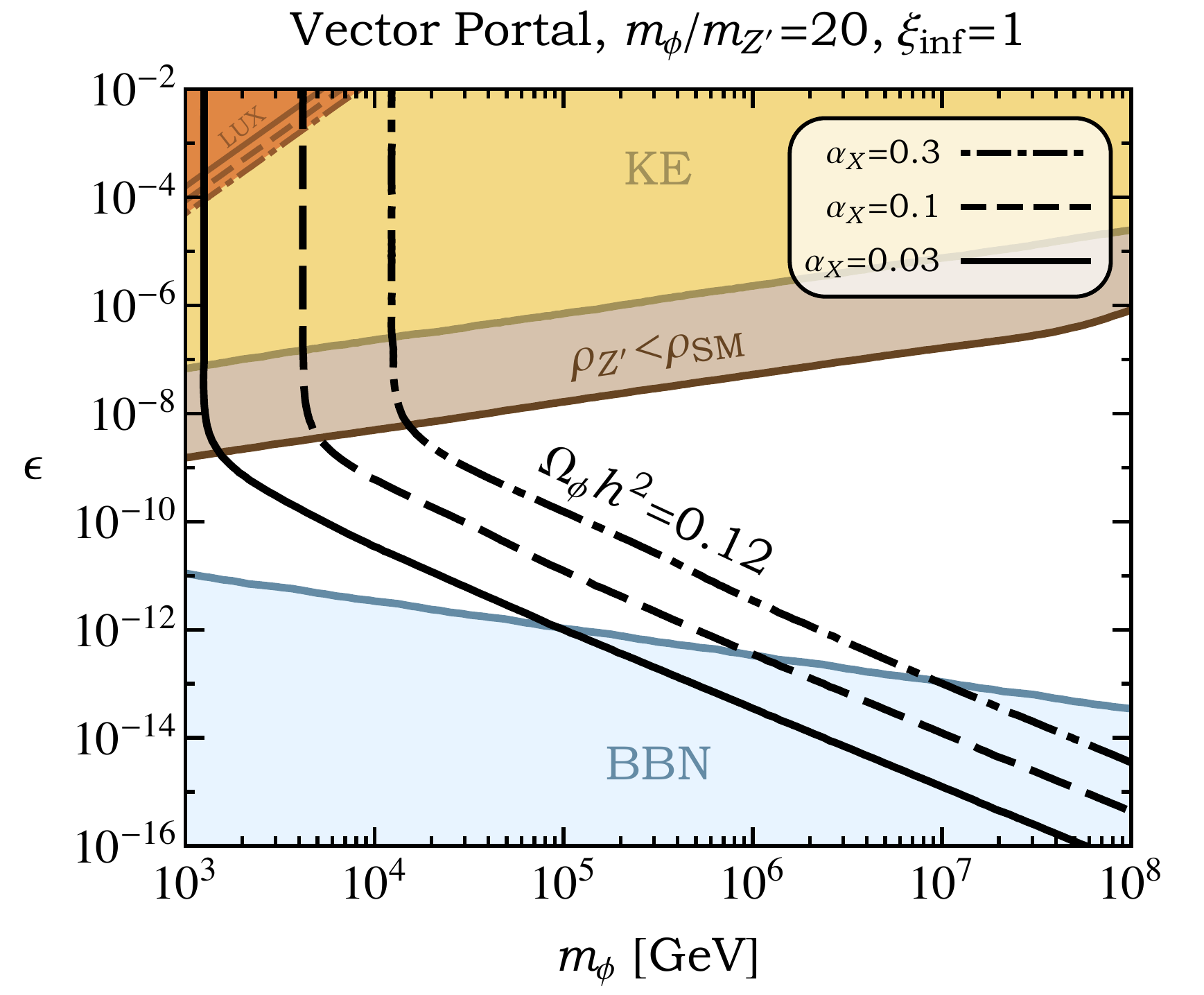}
\includegraphics[width=3.2in]{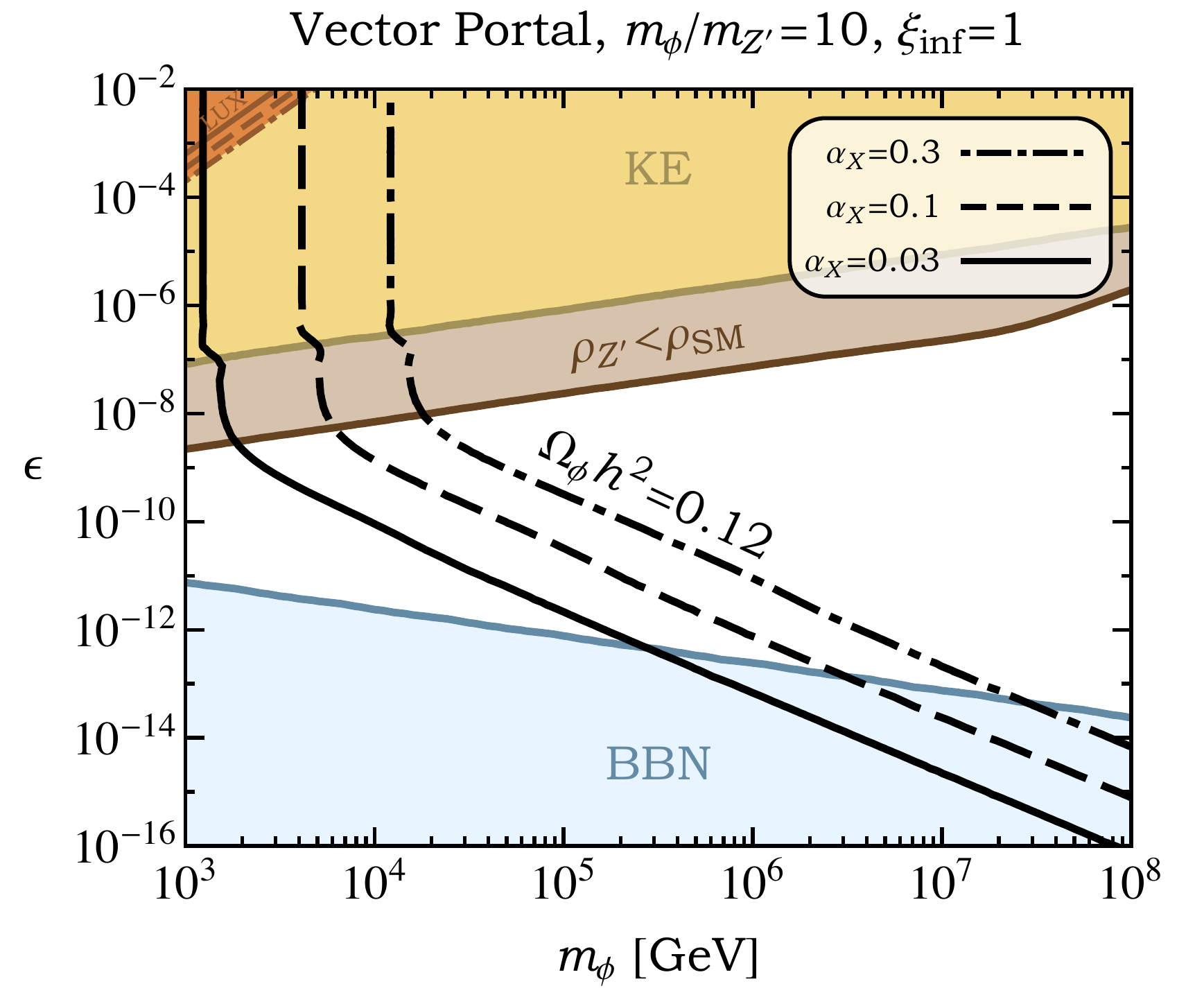}    \hspace{-0.7cm}    \\
  \hspace{-0.7cm} 
\includegraphics[width=3.2in]{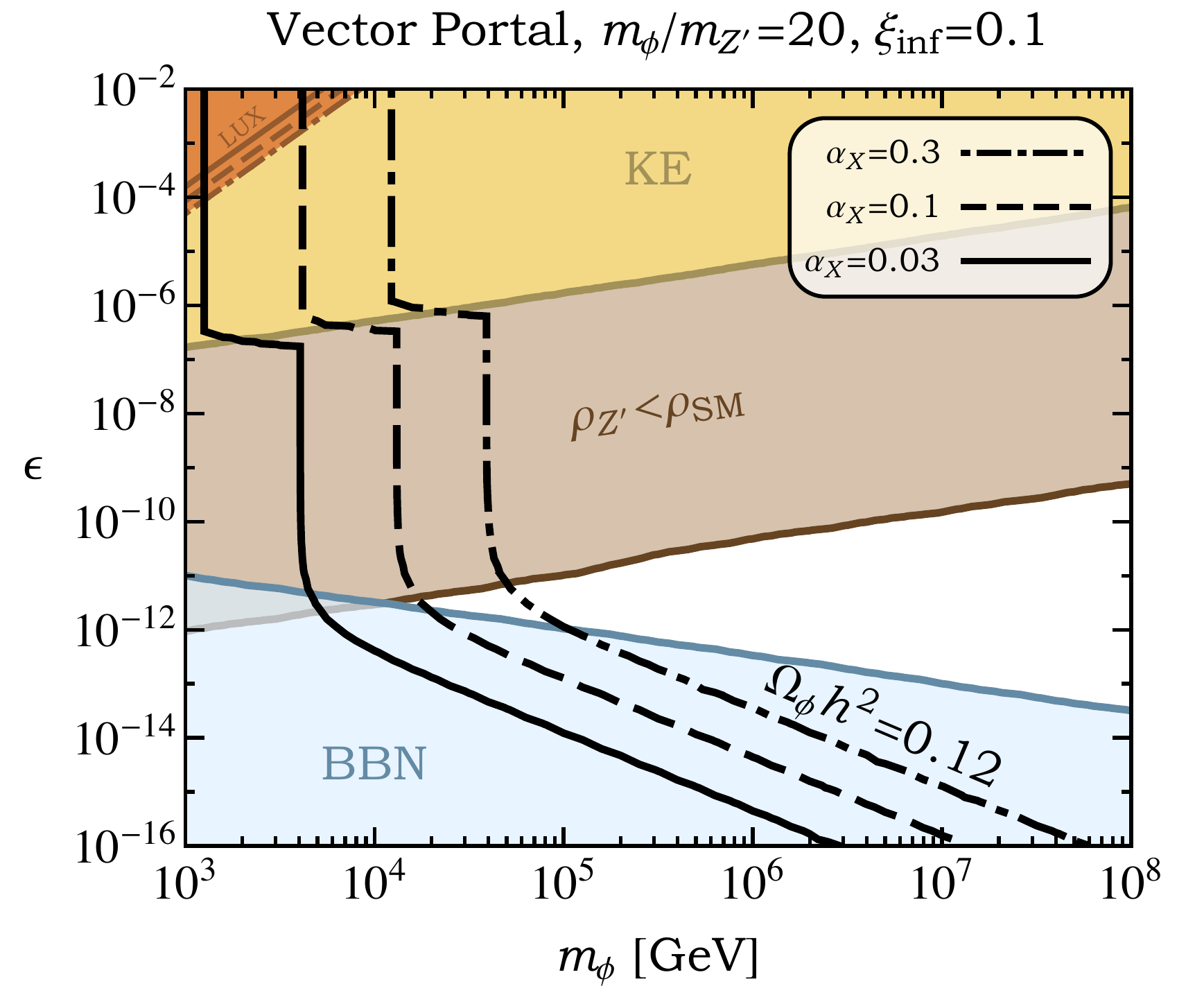} 
\includegraphics[width=3.2in]{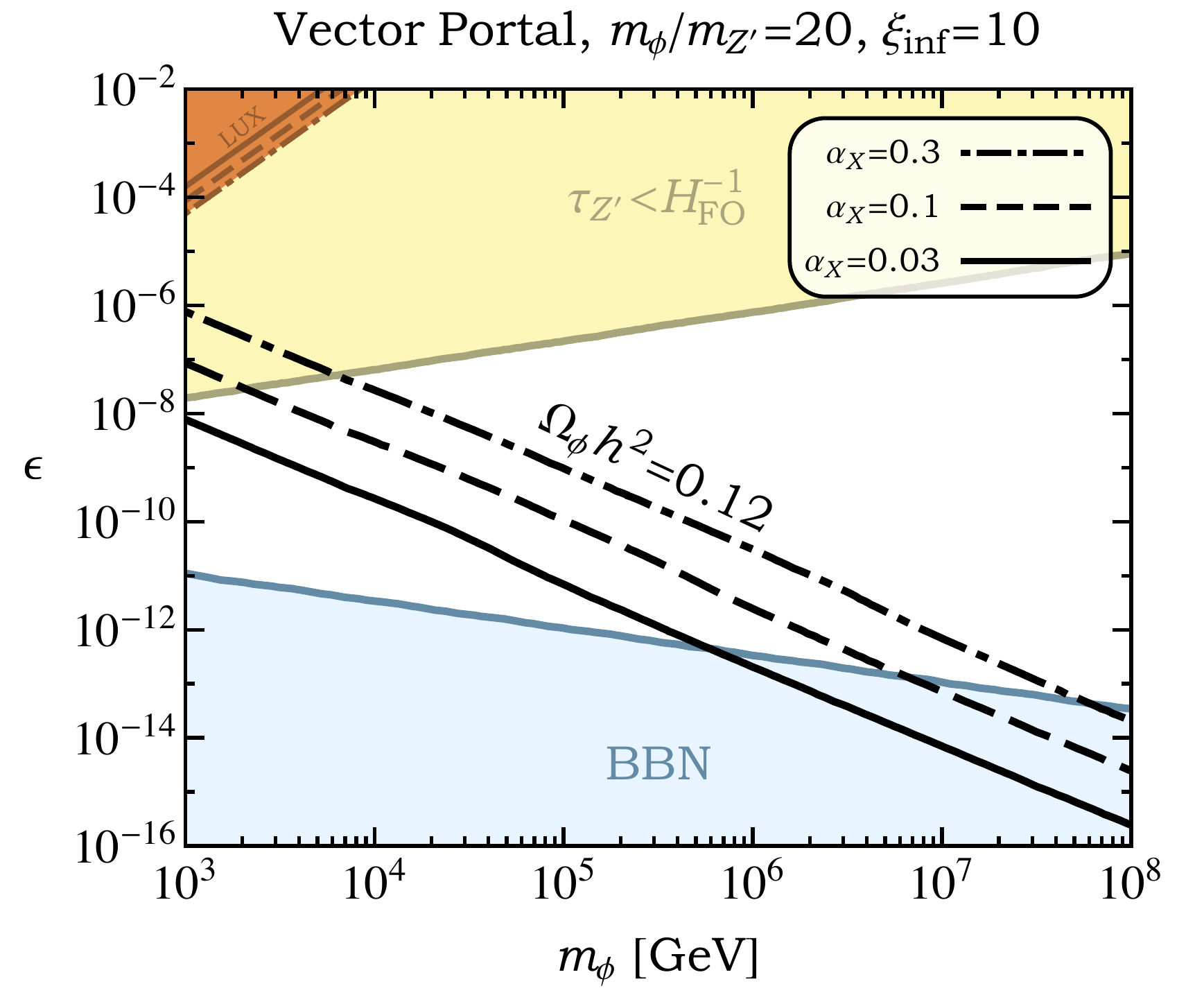}   
\caption{\label{fig:vector} Selected regions of parameter space in the vector portal model. The black contours ($\Omega_\phi h^2=0.12$) correspond to regions in the $m_\phi - \epsilon$ plane where the final $\phi$ abundance matches the observed dark matter density for three different values of the $Z^\prime$ coupling, $\alpha_X = 0.03$, 0.1, and 0.3. For larger values of $\epsilon$, and for the same three values of $\alpha_X$, the red regions (LUX) are currently ruled out by direct detection constraints from LUX and/or PandaX~\cite{Akerib:2015rjg,Tan:2016zwf}. On the other hand, in the shaded blue region (BBN) the $Z^\prime$ decays reheat the SM plasma to a temperature below 10 MeV, in potential tension with the successful predictions of BBN. In and above the brown region ($\rho_{Z^\prime} < \rho_\text{SM}$), the $Z^\prime$ population never comes to dominate the energy density of the universe, while in and above the yellow region ($\tau_{Z^\prime} < H_\text{FO}^{-1}$) $Z^\prime$ dominates the energy density but decays before the freeze-out of $\phi$. The shaded orange region (KE) corresponds to values of $\epsilon$ for which kinetic equilibrium between the hidden and visible sectors is established. In the top-left and top-right panels, we have fixed $\xi_\text{inf}=1$ and  $m_{\phi} / m_{Z^\prime}= 20$ and 10, respectively. The bottom-left and bottom-right panels illustrate the effect of varying $\xi_\text{inf}$ while fixing $m_{\phi} / m_{Z^\prime}= 20$.}
\end{center}
\end{figure}

We illustrate the phenomenology of this model in Fig.~\ref{fig:vector}, as a function of the dark matter mass, $m_\phi$, and kinetic mixing parameter, $\epsilon$, for various values of $\alpha_X$, $m_\phi / m_{Z^\prime}$, and $\xi_\text{inf}$. The abundance of dark matter in the vector portal scenario diverges from the typical WIMP estimate for sufficiently small values of $\epsilon$. In this case, $\phi$ can be as heavy as $\order{10}$ PeV before running afoul of constraints from Big Bang Nucleosynthesis (BBN). Along the black contours, the final abundance of $\phi$ matches the observed dark matter density, $\Omega_\phi h^2 \sim 0.12$. For longer lifetimes of the $Z^\prime$ (smaller values of $\epsilon$), $Z^\prime$ can come to dominate the energy density of the universe, corresponding to the parameter space below the brown shaded region. In this case, the entropy dump from the $Z^\prime$ decay significantly dilutes the $\phi$ abundance, allowing for large values of $m_\phi$ which would otherwise be inconsistent with the observed density of dark matter. For lifetimes longer than $\order{1}$ second, however, the reheating temperature after the $Z'$ decay is significantly less than 10 MeV, leading to potential tension with the successful predictions of BBN (shaded blue). For a sufficient degree of kinetic mixing, the hidden sector and SM bath are maintained in kinetic equilibrium in the early universe (shaded orange), and may potentially fall within the reach of direct detection experiments, such as LUX~\cite{Akerib:2016vxi,Akerib:2015rjg} and PandaX~\cite{Tan:2016zwf} (shaded red). We also highlight the parameter space in which the $Z^\prime$ population dominates the energy density and decays before the freeze-out of $\phi$ (shaded yellow). In this case, the hidden and visible sector entropies are no longer conserved during the freeze-out of $\phi$, invalidating the assumption that led to the derivation of Eqs.~(\ref{eq:boltz3}) and (\ref{eq:relicab}). Since, in most cases, the $Z^\prime$ abundance does not dominate the energy density of the universe when it decays before the freeze-out of $\phi$, we expect the resulting correction to be small. 

For $\xi_\text{inf} \ll 1$, as considered in the bottom-left panel of Fig.~\ref{fig:vector}, the hidden sector is only modestly populated (relative to the SM) after inflation. As a result, the effects of the $Z^\prime$ decay are reduced, and only for much longer lifetimes does the $Z^\prime$ population come to dominate the energy density of the universe. Regardless, compared to the standard thermal WIMP calculation, thermal decoupling in this scenario results in the underproduction of the hidden sector and thus allows for larger dark matter masses, without exceeding the observed dark matter density. 


Although we have focused on scalar dark matter in this section, fermonic dark matter is also a viable possibility within the context of vector portal scenarios~\cite{Berlin:2016vnh}. Qualitatively, very similar conclusions are reached in these two cases. In particular, Fig.~2 of Ref.~\cite{Berlin:2016vnh} can be directly compared to the results shown in Fig.~\ref{fig:vector} of this paper.

\subsection{Higgs Portal}
\label{sec:higgs}

In the Higgs portal scenario, a real scalar singlet, $\phi$, couples to the SM Higgs at tree-level~\cite{Pospelov:2007mp,Burgess:2000yq,Davoudiasl:2004be,Bird:2006jd,Kim:2006af,Finkbeiner:2007kk,D'Eramo:2007ga,Barger:2007im,SungCheon:2007nw,MarchRussell:2008yu,McDonald:2008up,Pospelov:2011yp,Piazza:2010ye,Kouvaris:2014uoa,Kainulainen:2015sva,Krnjaic:2015mbs,Batell:2012mj,Tenkanen:2016jic}. Working in the basis where $\phi$ does not acquire a VEV, the general scalar potential is given by
\be
\label{eq:FullScalarPotential}
V (\phi, H) = - \mu^2 |H|^2 + \lambda |H|^4 + \frac{\delta_1}{2} \, |H|^2 \, \phi + \frac{\delta_2}{2} \, |H|^2 \, \phi^2 - \frac{\delta_1 v^2}{4} \, \phi + \frac{\kappa_2}{2} \, \phi^2 + \frac{\kappa_3}{3!} \, \phi^3 + \frac{\kappa_4}{4!} \, \phi^4
~,
\ee
where $v \equiv \mu/\sqrt{\lambda} \approx 246 $ GeV and the tadpole coefficient is chosen to prevent $\phi$ from getting a VEV.
After electroweak symmetry breaking, mass mixing between the SM Higgs, $h$, and $\phi$ is controlled solely by the dimensionful parameter $\delta_1$. In the limit that $m_\phi \approx \sqrt{\kappa_2} \gg m_h$, the mixing angle, $\epsilon$, is approximated as
\be
\epsilon = -\frac{v \, \delta_1}{2 m_\phi^2} + \order{m_h / m_\phi}~,
\ee
so we will therefore focus on the phenomenology that arises from the following simplified scalar potential
\be
\label{eq:ScalarPotential}
V (\phi,H) = V_\text{SM} (H) + \frac{\delta_1}{2} \, |H|^2 \, \phi - \frac{\delta_1 \mu^2}{4} \, \phi + \frac{\kappa_2}{2} \, \phi^2 
~,
\ee
where $V_\text{SM} (H)$ is the SM Higgs potential. Note that it is technically natural for $\delta_1$ to be very small. In particular, the quantum correction to $\delta_1$ via a SM Higgs loop scales as $\Delta \delta_1 \sim \delta_1 \,  \lambda \, \log{(\Lambda_\text{UV}/m_{\phi})}/16\pi^2 \, $, where $\Lambda_\text{UV}$ is the high-energy cutoff of the theory. Since we will be most interested in regions of small mixing, $\epsilon \ll 1$, LHC constraints on SM Higgs couplings are negligible~\cite{Aad:2015gba,Khachatryan:2014jba}.


In this model, we assume that $\phi$ is odd under an approximate $Z_2$ symmetry, which is softly broken only by the super-renormalizable portal coupling, $\delta_1$, in the simplified potential of Eq.~(\ref{eq:ScalarPotential}). For sufficiently long $\phi$ lifetimes, corresponding to small values of $\epsilon$, we typically need $\delta_1$ to be in the neighborhood of  
\be
\delta_1 \simeq {\rm GeV}  \left(\frac{\epsilon}{10^{-10}}\right)\left(\frac{m_\phi}{\rm PeV}\right)^2~,~
\ee
in the general vicinity of the weak scale. In the full potential of Eq.~(\ref{eq:FullScalarPotential}), there is an additional $Z_2$ breaking coupling, $\kappa_3$, which renormalizes the value of $\delta_1$ at the one loop level. To ensure that this does not significantly increase the $\phi$ width, this correction must not exceed $ \sim \delta_1$, which implies
\be
\kappa_3 \lesssim \frac{16 \pi^2   }{ \log \frac{\Lambda_{\rm UV}}{m_\phi} }  \frac{\delta_1}{\delta_2}~.~
\ee


As our dark matter candidate, we introduce a singlet Majorana fermion, $\x \, $, which couples to $\phi$ through the interactions
\be
\mathcal{L} \supset ~ \phi ~ \bar{\x} (\lambda_s + \lambda_p \, i \gamma^5) \x 
~.
\ee
After EWSB, mass mixing leads to the substitution $\phi \to \phi - \epsilon ~ h$, which generates an effective dark matter coupling to the SM Higgs, allowing direct detection experiments to constrain the quantity $\lambda_s \, $. In particular, $\x$ scatters off nucleons through SM Higgs exchange with a spin-independent cross section of
\be
\sigma_\text{SI} \approx 2 \times 10^{-46} \text{ cm}^2 \times \left( \frac{\epsilon}{0.1} \right)^2 ~ \left( \frac{\lambda_s}{0.1} \right)^2
~.
\ee
Similarly, $\phi$ couples directly to the SM through
\be
\mathcal{L} \supset - \frac{\delta_1}{4} \, \phi \, h^2  - \, \frac{\epsilon}{v}\sum\limits_f m_f  ~ \phi \, \bar{f} f + \frac{2 \epsilon}{v} \left( m_W^2 \, W^{+ \mu} W_\mu^- + \frac{1}{2} m_Z^2 \, Z^\mu Z_\mu \right) \left( \phi + \frac{1}{v} \,  h \, \phi \right) + \order{\epsilon^2}
~.
\ee
At leading order in $\epsilon$, $\phi$ decays to pairs of Higgs bosons, SM fermions, and gauge bosons, with partial widths given by
\begin{align}
\Gamma (\phi \to h h) & = \frac{m_\phi^3 \epsilon^2}{32 \pi v^2} + \order{m_h / m_\phi}
\nl
\Gamma (\phi \to f \bar{f}) &= \frac{n_c m_\phi m_f^2 \epsilon^2}{8 \pi v^2} + \order{m_f / m_\phi}
\nl
\Gamma (\phi \to V V ) &= 
\frac{m_\phi^3 \epsilon^2}{16 \pi (1 + \delta_{V Z}) v^2}  + \order{m_V / m_\phi}
~,
\end{align}
where $n_c$ is the number of colors of the SM fermion, $f$, and $\delta_{V Z} = 1 (0)$ for $Z$ ($W^\pm$) final states.
As seen from the limiting forms above, $\Gamma (\phi \to h h) / \Gamma (\phi \to f \bar{f}) \sim (m_\phi / m_f)^2$ and $\Gamma (\phi \to h h) / \Gamma (\phi \to V V) \sim (m_\phi / m_V)^4$. Since we will focus here on cases in which $m_\phi \gg 100$ GeV, the dominant decay channel is to pairs of SM Higgs bosons. 

Prior to the decay of $\phi$, $\x$ freezes out through its annihilations within the hidden sector, $\x \x \to \phi \phi$. In particular, $\sigma v (\x \x \to \phi \phi) = a + b \, v^2$, where
\begin{align}
a &= \frac{2\sqrt{1-r^2} \lambda_p^2 \lambda_s^2}{m_\x^2 \pi  \left(r^2-2\right)^2} \approx \frac{\lambda_p^2 \lambda_s^2}{2 \pi  m_\x^2} + \mathcal{O}(r^2)
\nl
\nl
b &= \frac{-2 \left(r^2-1\right)^3 \lambda_p^4+3 \left(r^6-8 r^4+20 r^2-12\right) \lambda_s^2 \lambda_p^2+2 \left(-2 r^6+10 r^4-17 r^2+9\right) \lambda_s^4}{12 m_\x^2 \pi  \sqrt{1-r^2} \left(r^2-2\right)^4} 
\nl
&\approx \frac{\lambda_p^4-18 \lambda_p^2 \lambda_s^2+9 \lambda_s^4}{96 \pi  m_\x^2} + \mathcal{O}(r^2)
~,
\end{align}
where $v$ is the relative $\x$ velocity, and $r \equiv m_{\phi} / m_\x \, $. If $\phi$ departs from chemical equilibrium while relativistic, the initial abundance of $\x$ is given by Eq.~(\ref{eq:relicab}).

\begin{figure}[t]
\begin{center}
\includegraphics[width=0.497\textwidth]{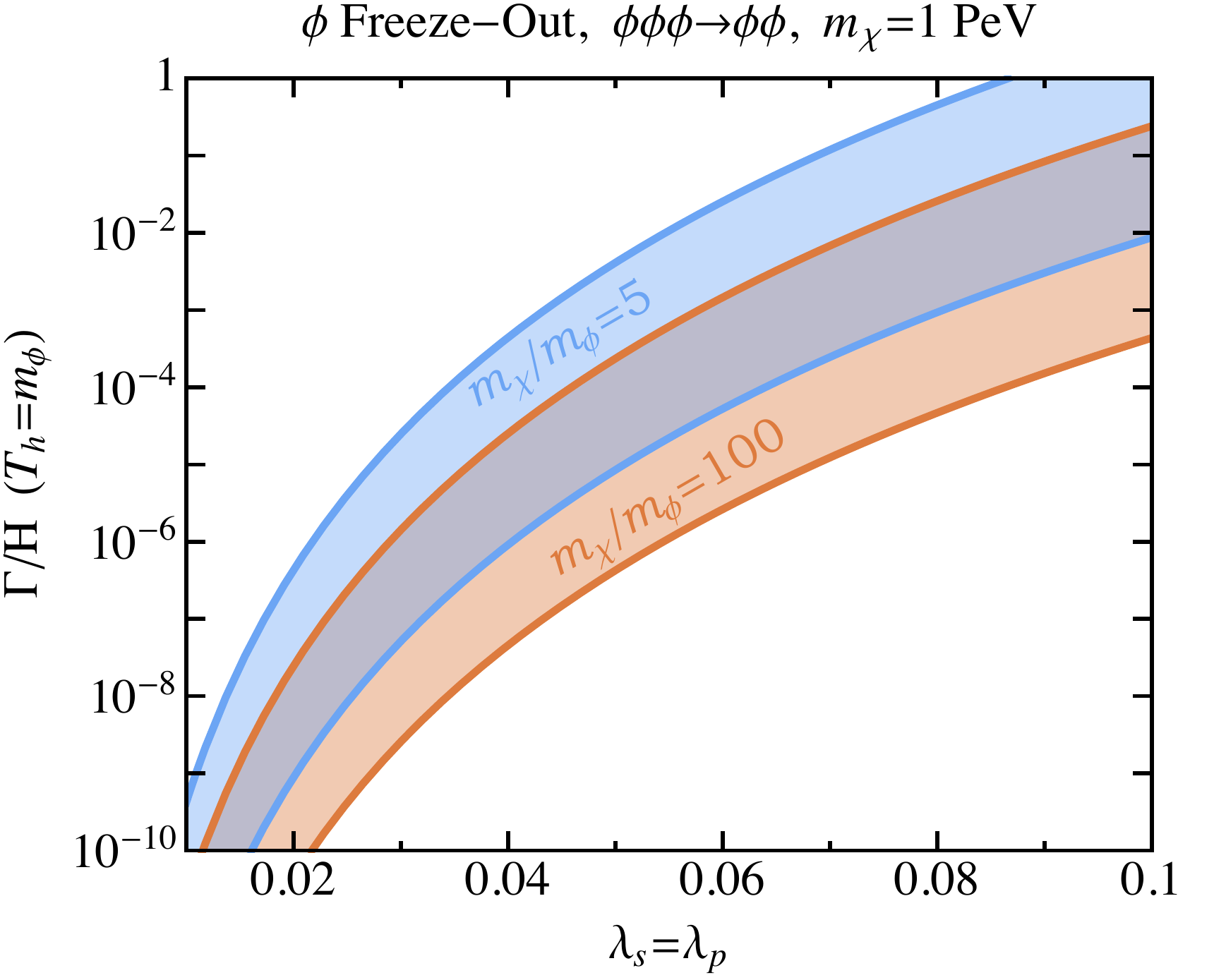}
\caption{\label{fig:Phi_FO} $\Gamma / H$ evaluated at $T_h = m_\phi$, as a function of $\lambda_s = \lambda_p$, for the process $\phi \phi \phi \to \phi \phi$, assuming that the hidden and visible sectors are thermally decoupled. We have taken $m_\x = 1$ PeV throughout and $m_\phi= 200$ TeV (blue) and 10 TeV (red). The width of the bands corresponds to $\xi_{\text{inf}}=0.1 - 10$. Larger values of $\xi_\text{inf}$ lead to larger rates relative to Hubble expansion. $\Gamma / H \lesssim 1$ only for $\lambda_{s,p} \lesssim \order{0.1}$. For larger values of $\lambda_{s,p}$, $\phi$ departs chemical equilibrium after becoming non-relativistic, and one must numerically solve the coupled Boltzmann equations for $\x$ and $\phi$.}
\end{center}
\end{figure}

For the case of the vector portal, as discussed in Sec.~\ref{sec:vector}, $Z^\prime$ depleting processes were suppressed. As we shall see below, however, $\phi$ is able to maintain chemical equilibrium in the Higgs portal case when $T_h \lesssim m_\phi$ for sufficiently large values of $\lambda_s$ or $\lambda_p\, $. This is directly tied to the fact that these interactions involve scalars and correspond to operators of lower dimension. Similar to as in the previous subsection, we consider the process $\phi \phi \phi \to \phi \phi$ mediated by a $\x$ loop. 
%
%
%
%
Following the approach described in Appendix~\ref{sec:app1}, we find by explicit calculation the rate for this process in the non-relativistic limit:
\be
\Gamma (\phi \phi \phi \to \phi \phi) =  n_\phi^2 ~ \frac{784 \sqrt{5}}{3 \pi^5} \frac{\lambda^{10}}{m_\x^2 m_\phi^3}
~,
\ee
where, for simplicity, we have taken $m_\x \gg m_\phi$ and $\lambda_s = \lambda_p = \lambda\, $. In Fig.~\ref{fig:Phi_FO}, we show $\Gamma (\phi \phi \phi \to \phi \phi) / H$ evaluated at $T_h = m_\phi$ as a function of $\lambda_s = \lambda_p\, $. It is apparent that if $\lambda_{s,p} \gtrsim \order{0.1}$, then $\Gamma / H \gtrsim 1$, indicating that $\phi$ freezes out while non-relativistic. In this case, instead of using Eq.~(\ref{eq:relicab}), we numerically solve the coupled Boltzmann system in Eq.~(\ref{eq:boltz3}).

If $\x$ and $\phi$ are to remain thermally decoupled from the SM during dark matter freeze-out, the scattering processes, $\phi h \leftrightarrow t \bar{t}$ and $\phi t \leftrightarrow h t$, must not exceed the rate of Hubble expansion before $T = m_\x / x_f$. At temperatures significantly above 100 GeV, the SM Higgs VEV, $v$, is suppressed, and hence, we will consider processes that do not depend explicitly on electroweak symmetry breaking, such as the $\phi-h-h$ cubic term in Eq.~(\ref{eq:ScalarPotential}) which is controlled by the dimensionful coupling, $\delta_1$. In the limit $m_\phi \gg m_t, m_h$, we find that the scattering processes are approximated as
\begin{align}
\sigma v (\phi \, h \to t \, \bar{t}) &\approx \frac{3 \delta_1^2 m_t^2}{32 \pi v^2 \, s (s - m_\phi^2)}~,
\nl
\sigma v (t \, \bar{t} \to \phi \, h) &\approx \frac{\delta_1^2 m_t^2 (s-m_\phi^2)}{128 \pi v^2 \, s^3}~,
\nl
\sigma v (\phi \, t \to h \, t) &\approx \frac{\delta_1^2 \, m_t^2}{64 \pi \, v^2 \, (s-m_\phi^2)} ~ \Bigg\{ ~ \frac{\left(4-x_h^2\right) s}{m_\phi^4 + x_h^2 \, s \, \left(s-m_\phi^2\right)} + \frac{1}{s-m_\phi^2} ~ \log{ \left[ \frac{s (s-m_\phi^2)^2}{m_t^2 \left(m_\phi^4 + x_h^2 s (s-m_\phi^2)\right)} \right] } ~ \Bigg\}~,
\nl
\sigma v (h \, t \to \phi \, t) &\approx \frac{\delta_1^2 \, m_t^2}{64 \pi \, v^2 \, s} ~ \Bigg\{ ~ \frac{\left(4-x_h^2\right) (s-m_\phi^2)}{m_\phi^4 + x_h^2 \, s \, \left(s-m_\phi^2\right)} + \frac{1}{s} ~ \log{ \left[ \frac{s (s-m_\phi^2)^2}{m_t^2 \left(m_\phi^4 + x_h^2 s (s-m_\phi^2)\right)} \right] } ~ \Bigg\}
~,
\end{align}
where $x_h \equiv m_h / m_t$, and the ``$v$" on the right-hand side denotes the SM Higgs VEV. If $n_{h,t} \, \sigma v \lesssim H$ at $T = m_\x / x_f$, then the hidden sector and the SM do not equilibrate before the freeze-out of the dark matter abundance. 

\begin{figure}[h!]
\begin{center} \hspace{-0.7cm} 
\includegraphics[width=3.2in]{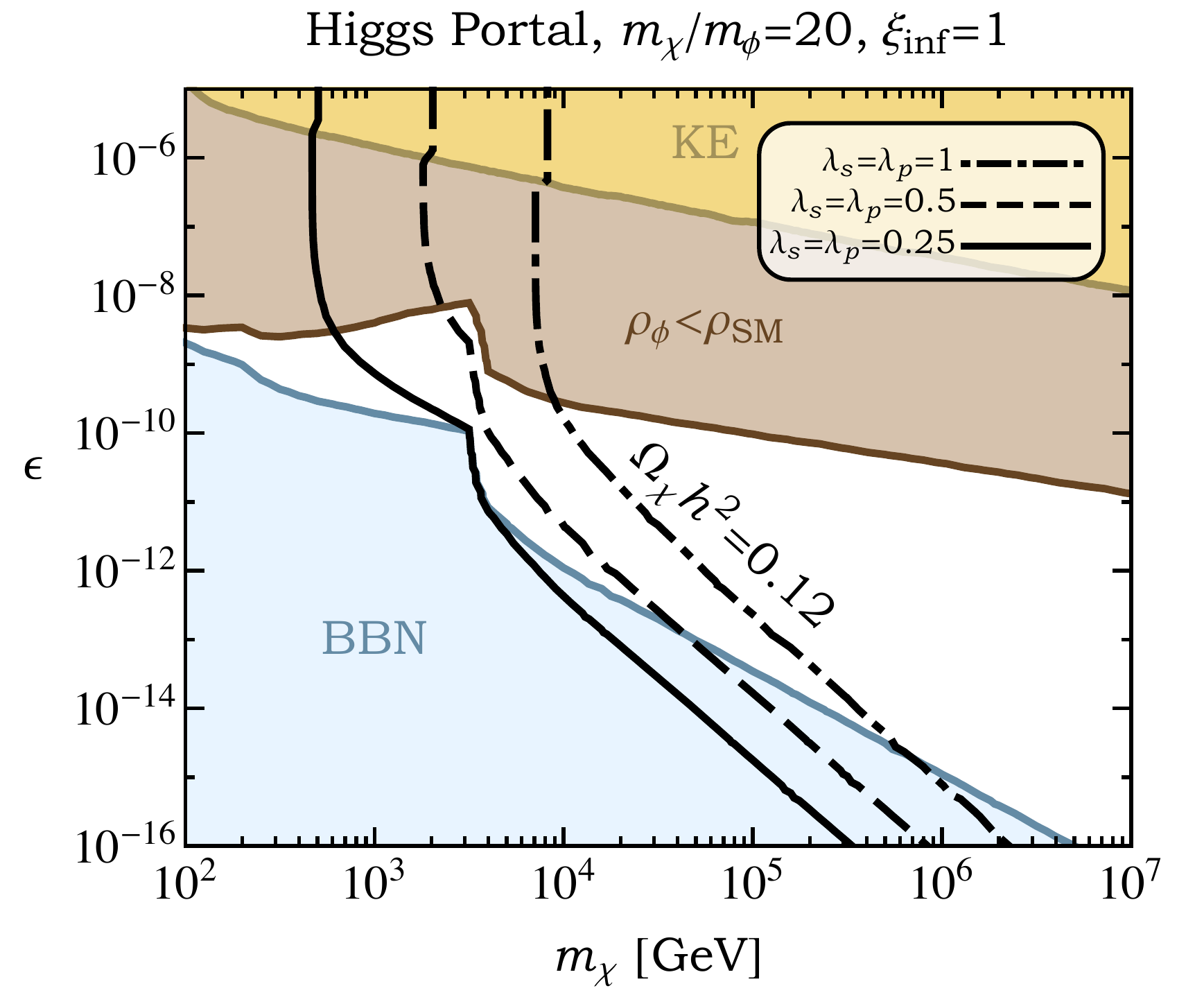}
\includegraphics[width=3.2in]{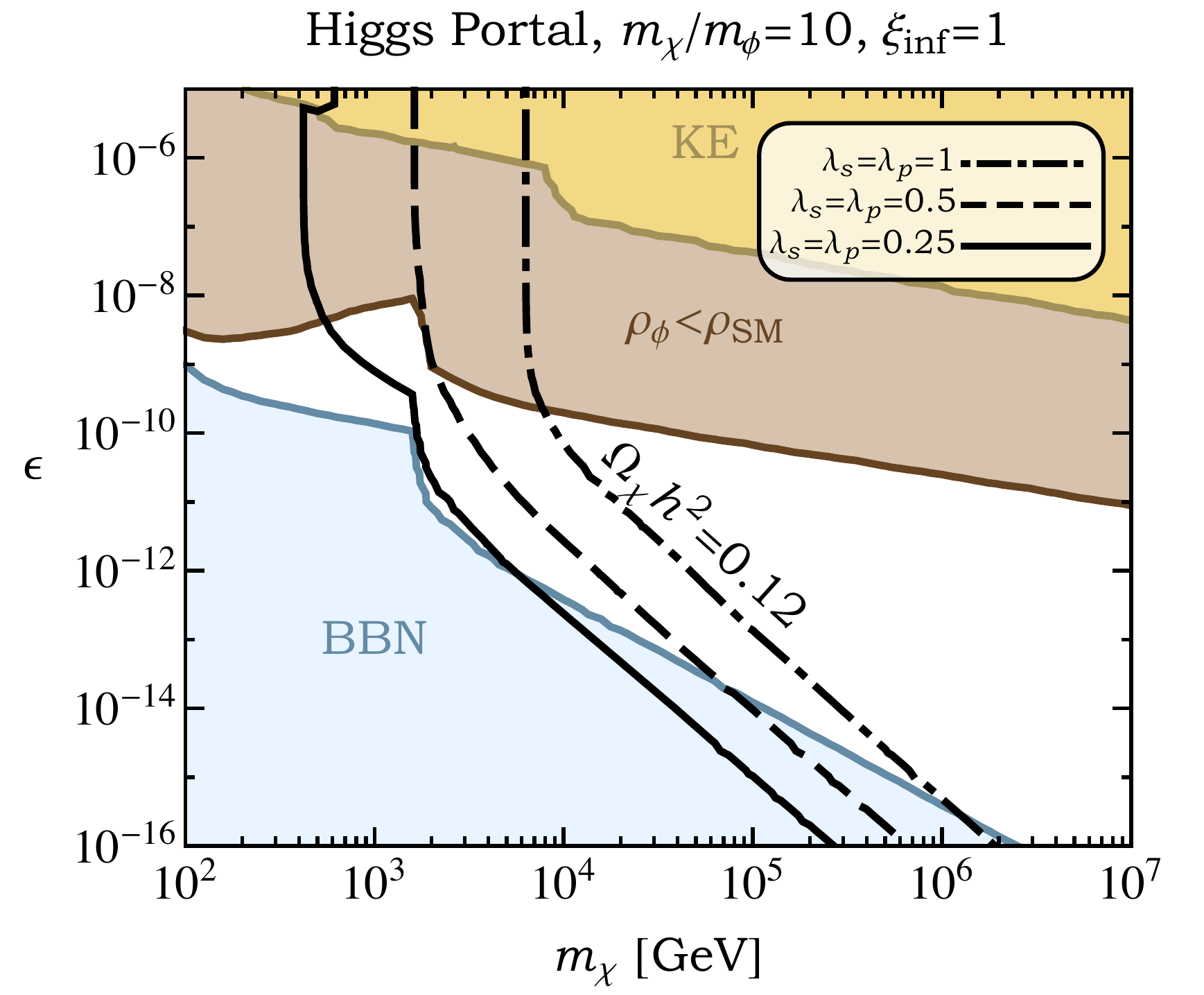} \\
\hspace{-0.7cm} 
\includegraphics[width=3.2in]{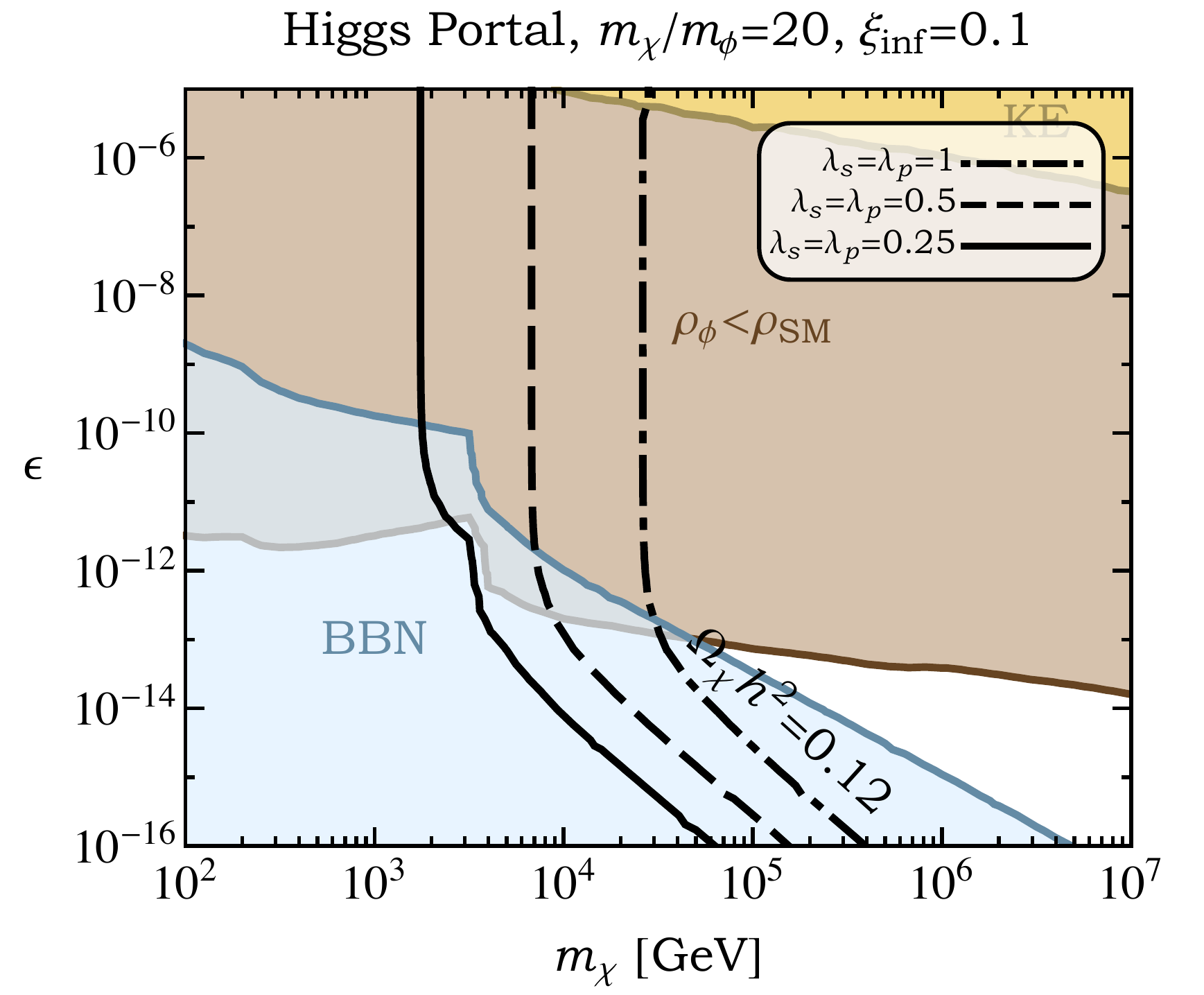}
\includegraphics[width=3.2in]{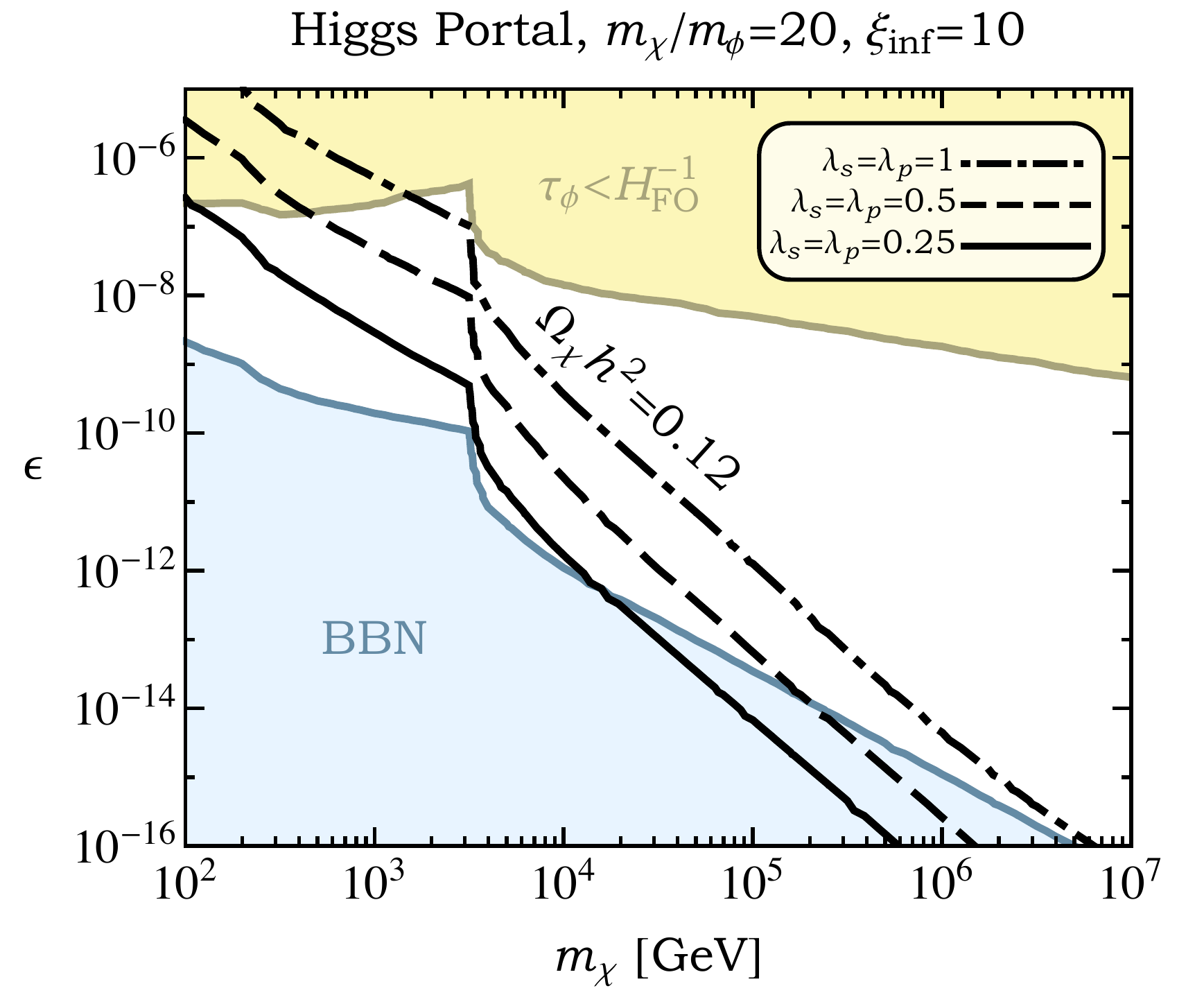}
\caption{\label{fig:higgs} Selected regions of parameter space in the Higgs portal model. The black contours ($\Omega_\phi h^2=0.12$) correspond to regions in the $m_\phi - \epsilon$ plane where the final $\phi$ abundance matches the observed dark matter density for three different values of the $\x-\phi$ couplings, $\lambda_s = \lambda_p = 0.25$, 0.5, and 1. The cross section for dark matter-nucleon scattering is beyond the reach of LUX or PandaX throughout the parameter space shown. In the shaded blue region (BBN) the $\phi$ decays reheat the SM plasma to a temperature below 10 MeV, in potential tension with the successful predictions of BBN. In and above the brown region ($\rho_\phi < \rho_\text{SM}$), the $\phi$ population never comes to dominate the energy density of the universe, while in and above the yellow region ($\tau_\phi < H_\text{FO}^{-1}$) $\phi$ dominates the energy density but decays before the freeze-out of $\x$. The shaded orange region (KE) corresponds to values of $\epsilon$ for which kinetic equilibrium between the hidden and visible sectors is established. In the top-left and top-right panels, we have fixed $\xi_\text{inf}=1$ and  $m_{\x} / m_{\phi}= 20$ and 10, respectively. The bottom-left and bottom-right panels illustrate the effect of varying $\xi_\text{inf}$ while fixing $m_{\x} / m_{\phi}= 20$. The jagged features depicted in some of these curves are the result of kinematic thresholds for $\phi$ decay, which predominantly proceed to heavy SM states.} 
\end{center}
\end{figure}

Similar to as in the previous section, Fig.~\ref{fig:higgs} illustrates the phenomenology of this model as a function of the dark matter mass, $m_\x$, and singlet-SM Higgs mixing parameter, $\epsilon$, for representative values of the quantities $m_\x / m_\phi$, $\xi_\text{inf}$ and $\lambda_{s,p}$. For simplicity, we consider the case that $\lambda_s = \lambda_p$. As discussed above, for $\lambda_{s,p} \gtrsim \order{0.1}$, the abundances of $\x$ and $\phi$ (prior to the decay of $\phi$) are calculated by numerically solving the coupled Boltzmann equations, Eq.~(\ref{eq:boltz3}), incorporating the dominant processes ($\x \x \to \phi \phi$ and $\phi \phi \phi \to \phi \phi$) that are responsible for the depletion of both species. Similar to as in Sec.~\ref{sec:vector}, for sufficiently suppressed values of the singlet-SM Higgs mixing parameter, $\epsilon$, $\phi$ is long-lived and comes to dominate the energy density of the universe, diluting the relic abundance of $\x$ upon its decay. However, compared to the vector portal scenario, this effect is suppressed, largely due to the enhanced strength of the process $\phi \phi \phi \to \phi \phi$. In particular, although larger values of $\lambda_{s,p}$ deplete the initial freeze-out abundance of $\x$ through the annihilations $\x \x \to \phi \phi$, such couplings also enhance the $3 \to 2$ self-annihilation for $\phi$, effectively depleting the comoving number density, $Y_\phi$, and softening the dilution from its decay, as seen from Eq.~(\ref{eq:suddendecay}). As a result, for $\lambda_{s,p} \sim \order{1}$, $\Omega_\x h^2$ matches the observed dark matter abundance without running afoul of constraints from BBN only for $m_\x \lesssim \order{100}$ TeV, when $\xi_\text{inf} = 1$ and $m_\x / m_\phi = 20$.

\subsection{Lepton Portal}
\label{sec:lepton}

The gauge singlet operator, $LH$, allows for the simple construction of a model that links the hidden and visible sectors through the lepton portal~\cite{Pospelov:2007mp,Bai:2014osa}. This same operator is often invoked in seesaw models as an explanation for the smallness of the SM neutrino masses~\cite{Minkowski:1977sc,Yanagida:1979as,Mohapatra:1979ia,GellMann:1980vs,Schechter:1980gr}. For realistic models of neutrino masses and mixing angles, there must be at least two right-handed SM singlet neutrinos, $N_{1,2}$, with Yukawa couplings to the SM lepton and Higgs doublets. As a result, models of neutrino masses often involve adding several new parameters to the SM Lagrangian, most of which are irrelevant to the dark matter phenomenology. Therefore, we will choose to focus on a simplified model involving only a single sterile neutrino, $N$, which couples to a single lepton doublet, $L$, where $L$ is one of the SM leptons, $L_e, L_\mu, L_\tau$~\cite{Batell:2016zod}. Additionally, as our dark matter candidate, we will add a SM singlet Weyl fermion, $\x$, and a real scalar, $\phi$, which will allow $\x$ to annihilate through the process $\x \x \to N N$. 

The relevant terms in the simplified Lagrangian take the form,
\be
- \mathcal{L} \supset  y_\nu \, N \, L \, H + \frac{1}{2} M_N \, N^2 + \lambda \, \phi \, \x \, N + \text{h.c.}
~,
\ee
where 2-component Weyl and $SU(2)_L$ indices are implied. For generality, and in light of the necessity of CP violation for leptogenesis, we will allow for $y_\nu$ and $M_N$ to be complex, but for simplicity take $\lambda$ to be real. In particular, the phases are parameterized as 
\be
y_\nu = |y_\nu| \, e^{i \phi_\nu} ~,\quad M_N = |M_N| \, e^{i \phi_N}
~. 
\ee
Although only one of these two phases is physical, we will allow for the presence of both explicitly in our calculation. We will also assume that $m_\phi > m_\x + |M_N|$ so that the $\phi$ decays promptly through $\phi \to \x N$ and hence does not repopulate the dark matter, $\x$, out of equilibrium. 

After electroweak symmetry breaking, the neutrino mass matrix is given by
\be
\label{eq:NeutrinoMassMatrix}
- \mathcal{L} \supset \frac{1}{2} \begin{pmatrix} \nu & N \end{pmatrix} \begin{pmatrix}  0 & y_\nu v / \sqrt{2} \\  y_\nu v / \sqrt{2} & M_N \end{pmatrix} \begin{pmatrix} \nu \\ N \end{pmatrix} + \text{h.c.}
~.
\ee
The physical masses are given by the square roots of the eigenvalues of $M^\star M$, where $M$ is the mass matrix in Eq.~(\ref{eq:NeutrinoMassMatrix}). In the limit that $|M_N| \gg |y_\nu| v$, the physical masses are
\be
m_{\nu_{_\text{SM}}} \approx \frac{ |y_\nu|^2 v^2 }{2 |M_N|} ~,\quad m_{\nu_s} \approx | M_N |
~.
\ee
The mass eigenstate basis is defined by, 
\be
\nu \approx - e^{i (\phi_N/2 - \phi_\nu)} \left( i ~\nsm - \epsilon ~ \nu_s \right) ~ , \quad N \approx e^{- i \phi_N/2 } \left(  \nu_s + i \, \epsilon ~\nsm \right)
~,
\ee
where
\be
\epsilon \equiv \frac{|y_\nu| \, v}{\sqrt{2} \, |M_N|} \approx \left(\frac{m_{\nu_{_\text{SM}}}}{m_{\nu_s}}\right)^{1/2}
~,
\ee
and $\nu_{_\text{SM}}$ and $\nu_s$ are predominantly SM-like and singlet-like, respectively.

%
%
%
%
Now, let us rewrite the relevant Lagrangian interactions (to leading order in $\epsilon$) in 4-component notation, taking into account all of the necessary field redefinitions to a basis in which $\nsm$, $\nu_s$, and $\x$ are now Majorana spinors. We find
\begin{align}
\label{eq:lepton_interactions}
\mathcal{L} &\supset - \frac{|y_\nu|}{\sqrt{2}} ~ h ~ \bar{\nu}_{_\text{SM}} i \gamma^5 \nu_s - \lambda ~ \phi ~ \bar{\nu}_s \left( \cos{\frac{\phi_N}{2}} + \sin{\frac{\phi_N}{2}} ~ i \gamma^5 \right) \x - \epsilon ~ \lambda ~ \phi ~ \bar{\nu}_{_\text{SM}} \left( \sin{\frac{\phi_N}{2}} - \cos{\frac{\phi_N}{2}} ~ i \gamma^5 \right) \x
\nl
& + \frac{\epsilon \, g_2}{2 c_w} ~ Z_\mu ~ \bar{\nu}_{_\text{SM}} i \gamma^\mu \nu_s  
+ \frac{\epsilon \, g_2}{2 \sqrt{2}} ~ \left[ e^{i (\phi_\nu - \phi_N / 2)}  ~ W_\mu^+ ~ \bar{\nu}_s \gamma^\mu (1- \gamma^5) \ell  + \text{h.c.} \right]
~.
\end{align}
For the moment, we will ignore aspects relevant to leptogenesis, e.g., CP violation, so that the phases $\phi_{\nu,N}$ are set to zero, and the Lagrangian takes a more simplified form
\begin{align}
\mathcal{L} &\supset - \frac{|y_\nu|}{\sqrt{2}} ~ h ~ \bar{\nu}_{_\text{SM}} i \gamma^5 \nu_s - \lambda ~ \phi ~ \bar{\nu}_s \, \x + \epsilon ~ \lambda ~ \phi ~ \bar{\nu}_{_\text{SM}} i \gamma^5 \x
\nl
& + \frac{\epsilon \, g_2}{2 c_w} ~ Z_\mu ~ \bar{\nu}_{_\text{SM}} i \gamma^\mu \nu_s  
+ \frac{\epsilon \, g_2}{2 \sqrt{2}} ~ \left[ W_\mu^+ ~ \bar{\nu}_s \gamma^\mu (1- \gamma^5) \ell  + \text{h.c.} \right]
~.
\end{align}

Before the decay of $\nu_s$, $\x$ freezes out via $\x \x \to \nu_s \nu_s$ through the $t$-channel exchange of $\phi$ with an initial abundance that is dictated by  $\sigma v (\x \x \to \nu_s \nu_s) = a + b v^2$, where
\begin{align}
a &=  \frac{\lambda ^2 m_\x^2 (r+1)^2 \sqrt{1-r^2}}{16 \pi  \left(m_\x^2 \left(r^2-1\right)-m_\phi^2\right)^2}
\nl
&\approx \frac{\lambda ^2 m_\x^2}{16 \pi  \left(m_\x^2+m_\phi^2\right)^2} + \mathcal{O}(r^2)~,
\nl
\nl
b &= \frac{\lambda ^2 m_\x^2 (r+1)^{3/2} \left(m_\x^4 \left(r^2-1\right)^2 (r (23 r-8)+4)-2 m_\x^2 m_\phi^2 (r-1) (r+1) (r (23 r-24)-8)+m_\phi^4 (r (23 r-40)+20)\right)}{384 \pi  \sqrt{1-r} \left(m_\phi^2-m_\x^2 \left(r^2-1\right)\right)^4}
\nl
&\approx \frac{\lambda ^2 m_\x^2 \left(m_\x^4-4 m_\x^2 m_\phi^2+5 m_\phi^4\right)}{96 \pi  \left(m_\x^2+m_\phi^2\right)^4} + \mathcal{O}(r^2)
~,
\end{align}
$v$ is the relative $\x$ velocity, and $r \equiv m_{\nu_s} / m_\x$. If $\nu_s$ departs from chemical equilibrium while still relativistic, the initial abundance of $\x$ is well approximated by the semi-analytic form in Eq.~(\ref{eq:relicab}).

The interactions in Eq.~(\ref{eq:lepton_interactions}) allow $\nu_s$ to decay to electroweak Higgs/gauge bosons and SM leptons. To leading order in $m_h / m_{\nu_s}$, the corresponding rates are given by
\begin{align}
\Gamma (\nu_s \to h ~\nsm) &\approx \Gamma (\nu_s \to Z ~\nsm) \approx \Gamma (\nu_s \to W^\pm \ell^\pm) \approx \frac{ \epsilon^2 m_{\nu_s}^3}{16 \pi v^2}
~.
\end{align}
Hence, in the case that $m_{\nu_s} \gg m_h$, the total width is approximated as
\be
\Gamma_{\nu_s} \approx  \frac{ 3 \epsilon^2 m_{\nu_s}^3}{16 \pi v^2}
~.
\ee
The elastic scattering of $\x$ with nuclei proceeds through loops involving $\phi$ and $\nu_s$ at leading order, resulting in rates that are well below the irreducible neutrino background.

\begin{figure}[t]
\begin{center}
\includegraphics[width=0.497\textwidth]{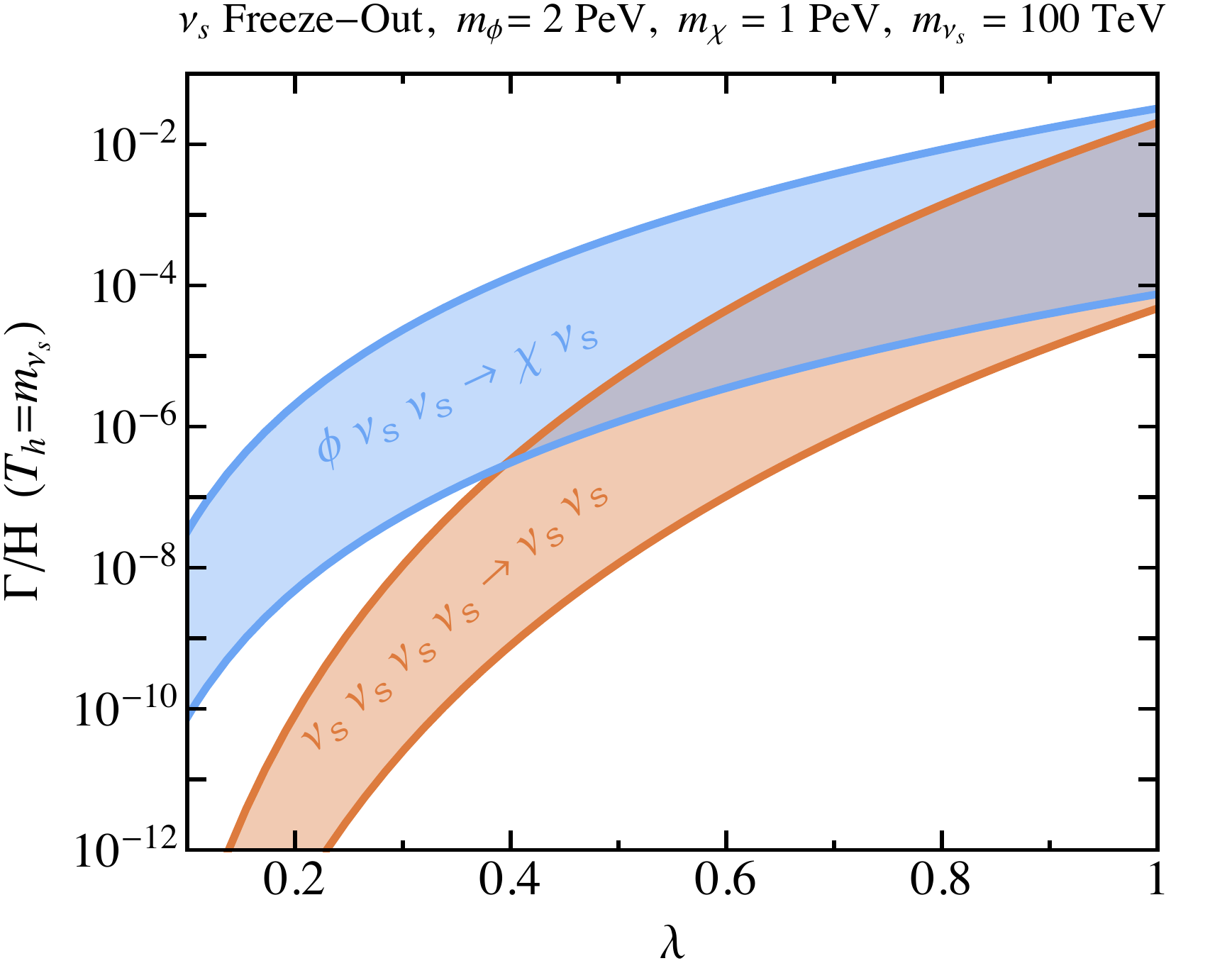}
\caption{\label{fig:Nu_FO} $\Gamma / H$ evaluated at $T_h = m_{\nu_s}$ as a function of the coupling $\lambda$ for the processes $\nu_s \nu_s \nu_s \nu_s \to \nu_s \nu_s$ (red), and $\phi \nu_s \nu_s \to \x \nu_s$ (blue), assuming that the hidden and visible sectors are thermally decoupled. We have taken $m_\phi = 2$ PeV, and $m_{\x}= 1$ PeV, and $m_{\nu_s} = 100$ TeV. The width of the bands corresponds to $\xi_{\text{inf}}=0.1 - 10$. Larger values of $\xi_\text{inf}$ lead to larger rates relative to Hubble expansion. For $\lambda \lesssim \order{1}$, $\Gamma / H \ll 1$ and hence $\nu_s$ departs chemical equilibrium while it is still relativistic.}
\end{center}
\end{figure}

After the freeze-out of $\x$, $\nu_s$ remains in chemical equilibrium until the rate for processes that deplete its number density falls below that of Hubble expansion. The process $\nu_s \nu_s \nu_s \nu_s \to \nu_s \nu_s$ is mediated by a $\x-\phi$ loop, similar to the left-most diagram of Fig.~\ref{fig:Zp_Feynman}. Assuming that $m_\phi \gg m_\x$, the rate for this process scales as 
\be
\Gamma (\nu_s \nu_s \nu_s \nu_s \to \nu_s \nu_s) \sim n_{\nu_s}^3 ~ \frac{\lambda^{12} \, m_{\nu_s}^2}{m_\phi^{10}}
~.
\ee
Similarly, $\phi \nu_s \nu_s \to \x \nu_s$ may proceed, e.g., through an $s$-channel $\phi$, analogous to the center and right-most diagrams of Fig.~\ref{fig:Zp_Feynman}. By dimensional analysis, we estimate the corresponding rate as
\be
\Gamma (\phi \nu_s \nu_s \to \x \nu_s) = n_{\nu_s} n_\phi  ~ \frac{\lambda^6}{m_\phi^5}
~.
\ee
In Fig.~\ref{fig:Nu_FO}, we plot the quantity $\Gamma / H$, evaluated at $T_h = m_{\nu_s}$, as a function of $\lambda$ for these two processes. As illustrated in this figure, for $\lambda \lesssim \order{1}$, $\Gamma / H \lesssim 10^{-2}$, and $\nu_s$ is not maintained in chemical equilibrium. For the remainder of our analysis, we will therefore assume that $\nu_s$ freezes out while relativistic. Following the discussion above Eq.~(\ref{eq:delta}), this implies that the $\nu_s$ comoving number density is fixed as  $Y_{\nu_s} \approx 0.01 ~ \xi_\text{inf}^3 \, $ and, as in Sec.~\ref{sec:vector}, this justifies calculating the initial freeze-out abundance of $\x$ through the use of the semi-analytic form in Eq.~(\ref{eq:relicab}).

\begin{figure}[!t]
\begin{center}
\includegraphics[width=3.2in]{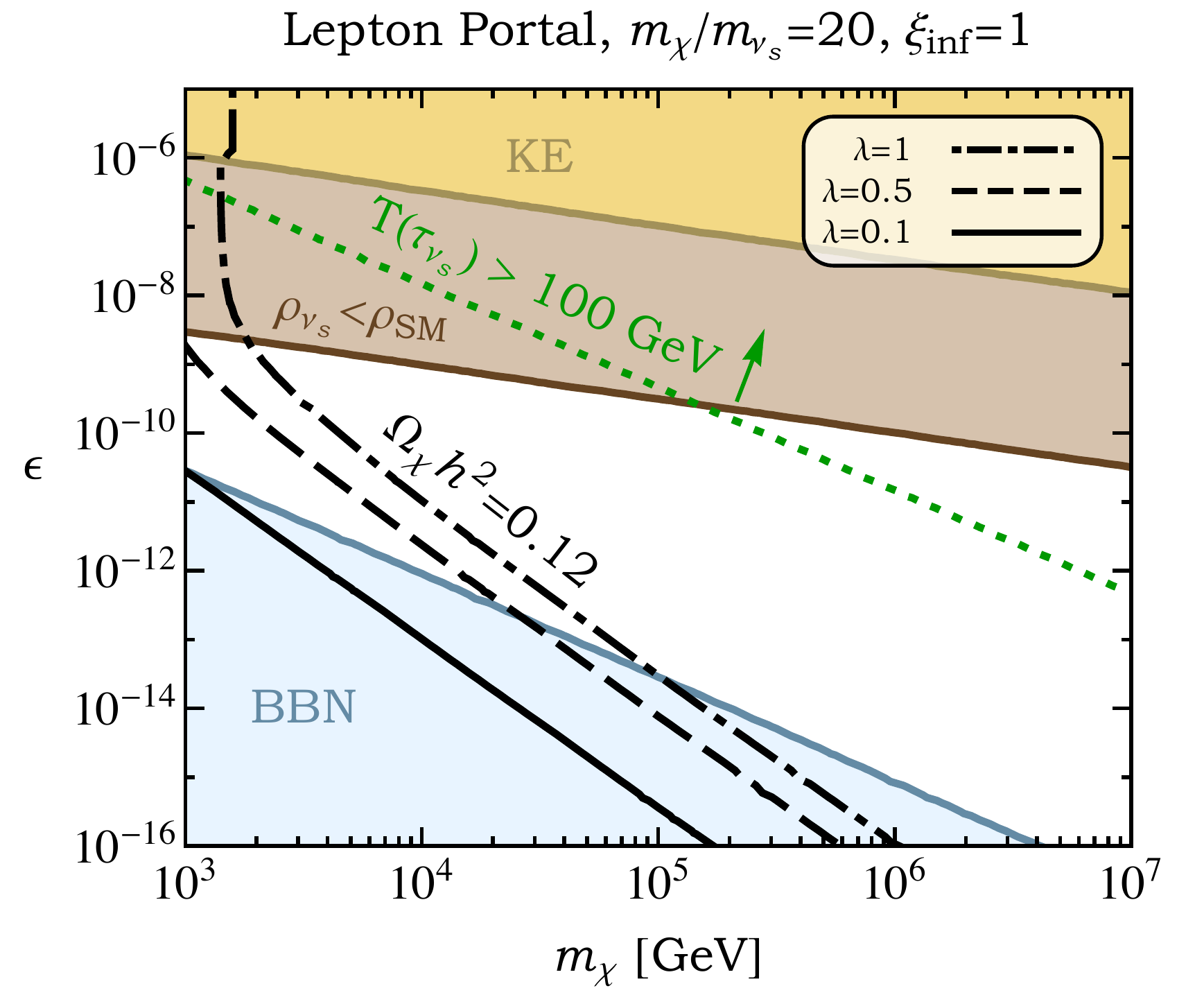}
\includegraphics[width=3.2in]{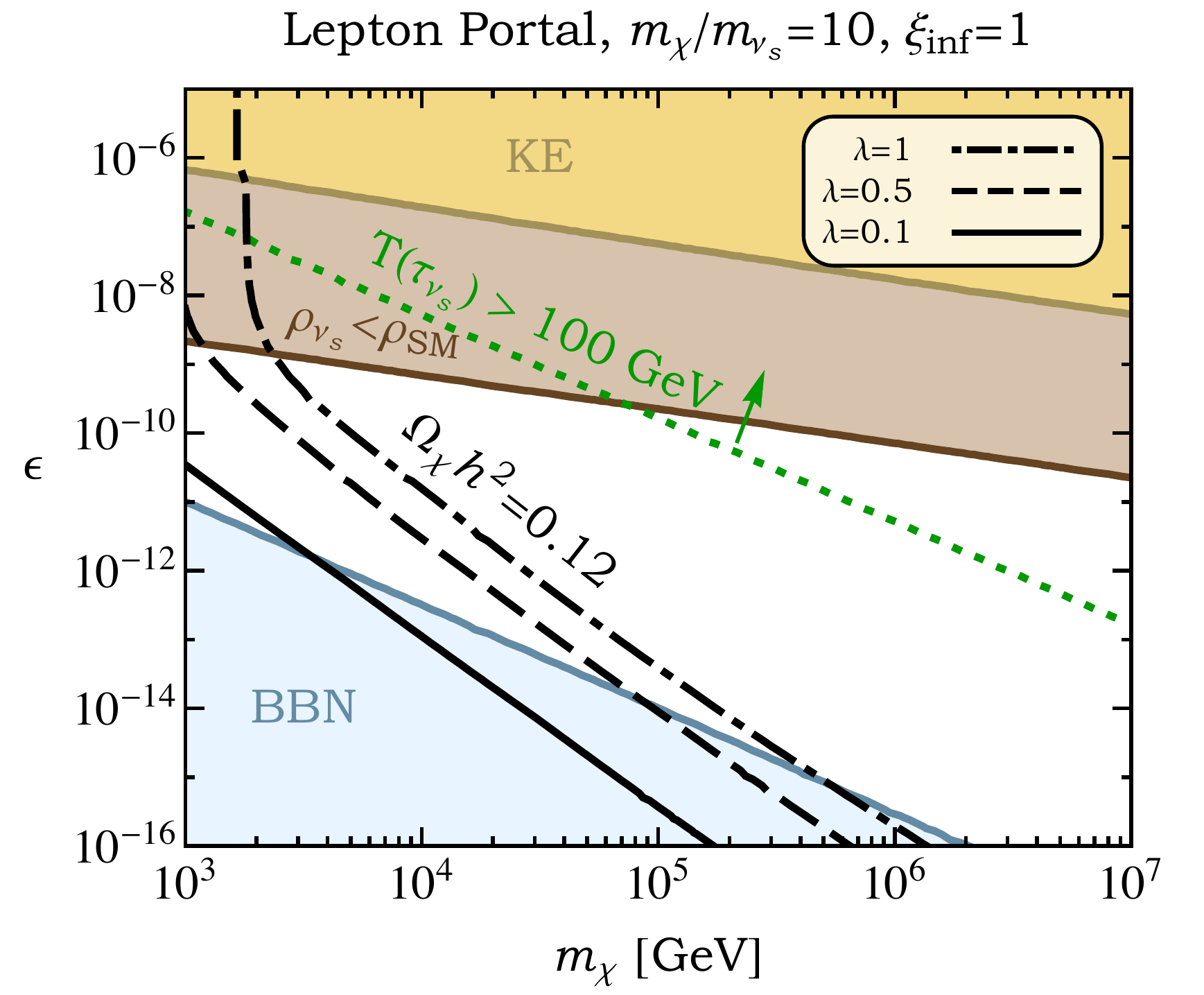}
\includegraphics[width=3.2in]{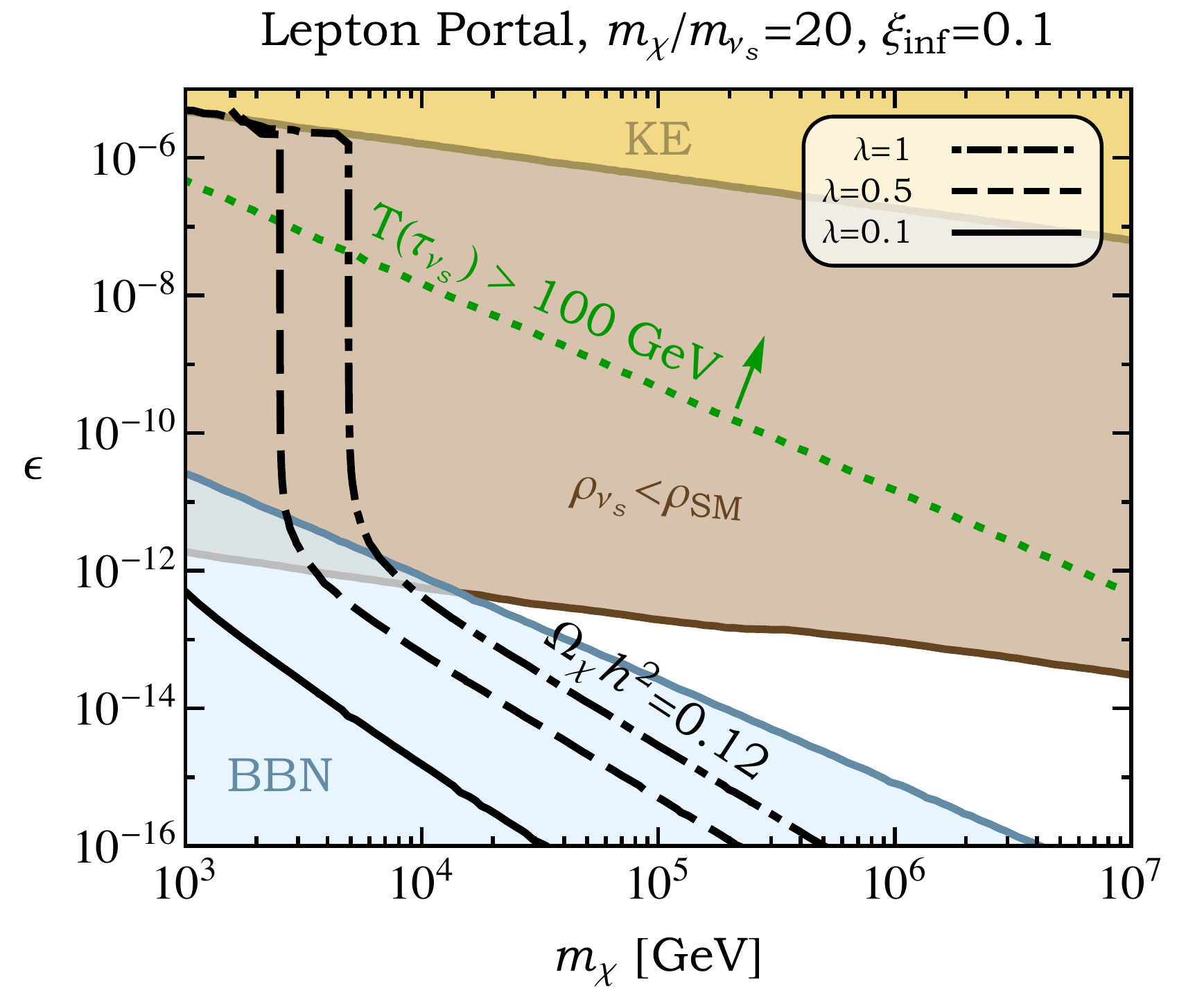}
\includegraphics[width=3.2in]{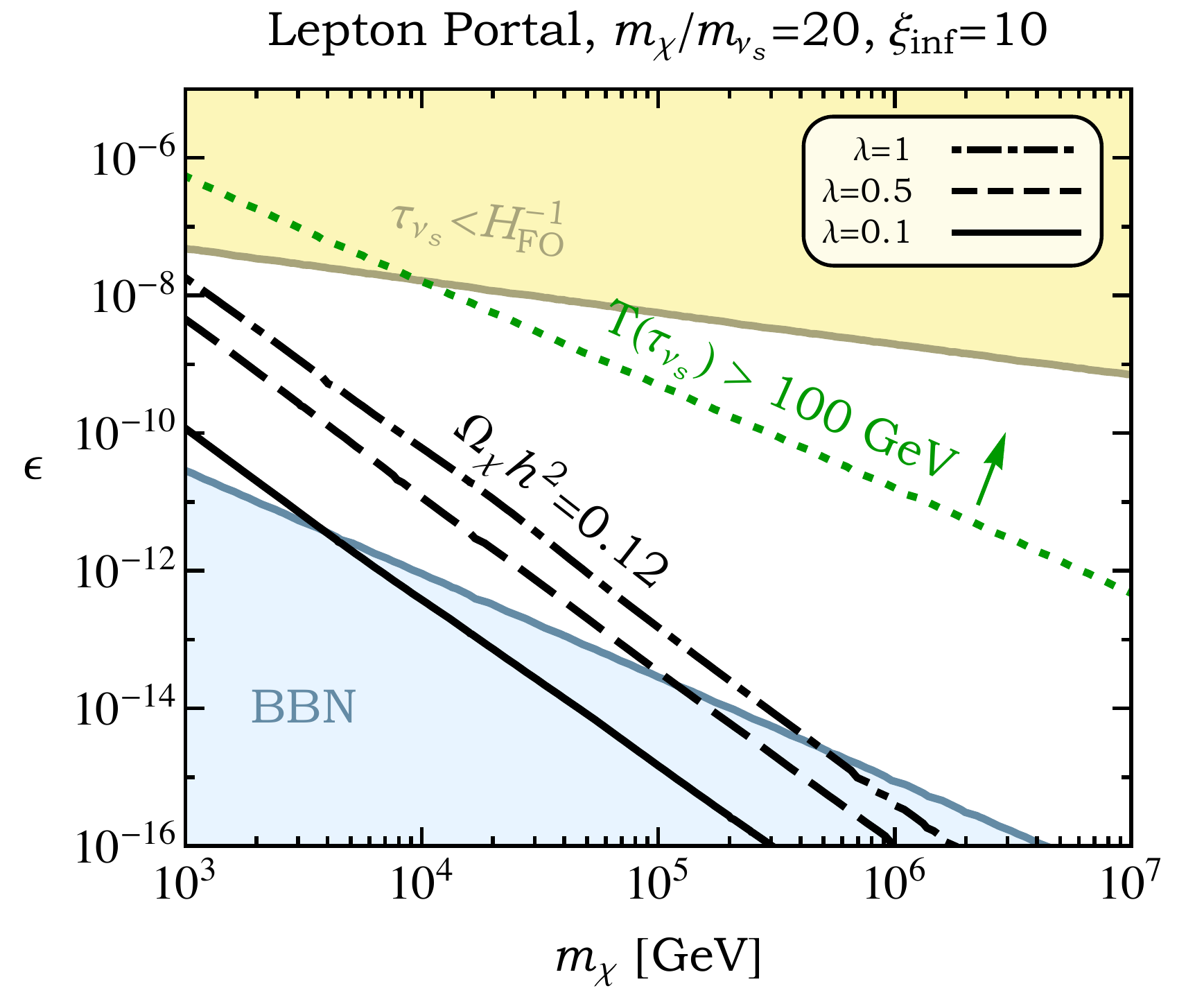}
\caption{\label{fig:lepton}
Selected regions of parameter space in the lepton portal model. The black contours ($\Omega_\x h^2 = 0.12$) correspond to regions in the $m_\x - \epsilon$ plane where the final $\x$ abundance matches the observed dark matter energy density for three different values of the $\x-\nu_s$ coupling, $\lambda = 0.1$, 0.5, and 1. The cross section for dark matter-nucleon scattering is beyond the reach of LUX or PandaX throughout the parameter space shown. In the shaded blue region (BBN) the $\nu_s$ decays reheat the SM plasma to a temperature below 10 MeV, in potential tension with the successful predictions of BBN. In and above the brown region ($\rho_{\nu_s} < \rho_\text{SM}$), the $\nu_s$ population never comes to dominate the energy density of the universe, while in and above the yellow region ($\tau_{\nu_s} < H_\text{FO}^{-1}$) $\nu_s$ dominates the energy density but decays before the freeze-out of $\x$. The shaded orange region (KE) corresponds to values of $\epsilon$ for which kinetic equilibrium between the hidden and visible sectors is established. The green dotted line ($T(\tau_{\nu_s}) > 100$ GeV) corresponds to the boundary of the parameter space in which the temperature of the SM plasma is above 100 GeV at the time of $\nu_s$ decay, representing a favorable condition for leptogenesis. In the top-left and top-right panels, we have fixed $\xi_\text{inf}=1$ and  $m_{\x} / m_{\nu_s}= 20$ and 10, respectively, while the bottom-left and bottom-right panels illustrate the effect of varying $\xi_\text{inf}$ while fixing $m_{\x} / m_{\nu_s}= 20$. The mass ratio, $m_{\phi}/m_{\x} = 1.1$, is fixed throughout all panels. 
}
\end{center}
\end{figure}

The hidden sector will remain thermally decoupled from the SM during the dark matter freeze-out process if the scattering processes $\nu_s\nsm \leftrightarrow t \bar{t}$ and $\nu_s t \leftrightarrow\nsm t$ do not exceed Hubble expansion before $T = m_\x / x_f$. For temperatures significantly above the electroweak scale, $\nu_s -\nsm$ mixing is suppressed, and hence, we will refrain from considering processes that depend explicitly on such mixing. In the limit that $m_{\nu_s} \gg m_t$, we find that the scattering cross sections for $\nu_s\nsm \leftrightarrow t \bar{t}$ and $\nu_s t \leftrightarrow\nsm t$ are approximated as
\begin{align}
\sigma v (\nu_s \nu_{_\text{SM}} \to  t \bar{t}) \approx \frac{3 |y_\nu|^2 m_t^2 s}{32 \pi v^2 (s-m_h^2)^2}~,~~
\sigma v (t \bar{t} \to \nu_s \nu_{_\text{SM}}) \approx \frac{|y_\nu|^2 m_t^2 (s-m_{\nu_s}^2)^2}{32 \pi v^2 s (s-m_h^2)^2}~, \nonumber
\end{align}
\begin{align}
\hspace{-0.8cm}\sigma v (\nu_s t \to \nsm t) &\approx \frac{m_t^2 |y_\nu|^2}{32 \pi v^2 (s-m_{\nu_s}^2)} \Bigg\{ \frac{m_{\nu_s}^4-2 m_{\nu_s}^2 s \left(x_h^2-2\right)+s^2 x_h^2}{m_{\nu_s}^4+s x_h^2 \left(s-m_{\nu_s}^2\right)} + \frac{m_{\nu_s}^2}{s-m_{\nu_s}^2} \log \left[ \frac{s \left(s-m_{\nu_s}^2\right)^2}{m_t^2 \left(m_{\nu_s}^4+s x_h^2 (s-m_{\nu_s}^2)\right)} \right] \Bigg\}~,  \!\!\!\!\!\!\!\!\!\!\!\!\! \nonumber
\end{align}
\begin{align}
\sigma v (\nsm t \to \nu_s t) \approx \frac{(s-m_{\nu_s}^2)^2}{s^2} \sigma v (\nu_s t \to \nsm t)~, 
\end{align}
where $x_h \equiv m_h / m_t$, and the ``$v$" on the right-hand side is the SM Higgs VEV. If $n_{\nsm} \, \sigma v \lesssim H$ at $T = m_\x / x_f$, then the hidden sector and the SM do not equilibrate before the freeze-out of the dark matter abundance.

In Fig.~\ref{fig:lepton}, we plot some of the phenomenological features of this model as a function of $m_\x$ and $\epsilon$, fixing $m_\phi = 1.1 ~ m_\x$ and for various choices of $m_\x / m_{\nu_S}$ and $\xi_\text{inf}$. In most respects, this resembles the results shown in the previous two subsections. In this case, however, we also show as a green dotted line the boundary of the region in which the temperature of the SM plasma is reheated to above 100 GeV through $\nu_s$ decays. Above this approximate temperature, electroweak sphalerons are in thermal equilibrium with the SM plasma, and are thus potentially able to convert a lepton-antilepton asymmetry (such as one generated through $\nu_s$ decays) into a baryon asymmetry.


\section{Summary and Conclusions}
\label{sec:conclusion}

Motivated by the increasingly stringent constraints that have been placed in recent years on dark matter in the form of WIMPs, we consider in this study dark matter candidates that are part of a larger sector with no sizable interactions with the Standard Model. Such a hidden sector could very plausibly be populated after inflation, and will undergo a thermodynamic history that is independent of the visible sector (which contains the Standard Model). As the hidden sector cools, its lightest particles will become non-relativistic and may come to dominate the energy density of the universe. When these particles ultimately decay, they reheat the universe and dilute the abundances of any previously frozen-out relics, including that of the dark matter itself. This sequence of events is a generic consequence of the hidden sector's highly decoupled nature, and phenomenology of this type can be found within a wide range of theoretical frameworks. 

In this study, we have described in some detail the thermodynamics and cosmological evolution of models that feature a highly decoupled hidden sector. After presenting a more general discussion, we have considered three simple, representative models, in which the hidden and visible sectors interact through what are known as the vector, Higgs, and lepton portals. In each of these cases, we identify significant parameter space in which the decoupled cosmological history considered here is viably realized. Furthermore, due to the dilution that results from the decays of long-lived hidden sector particles, the dark matter can be as heavy as $\sim$1-100 PeV in these scenarios, without generating a dark matter abundance in excess of the measured value.  

\bigskip
\bigskip
 
\textbf{Acknowledgments.} AB is supported by the Kavli Institute for cosmological physics at the University of Chicago through grant NSF PHY-1125897. Fermilab is operated by Fermi Research Alliance, LLC, under Contract No. DE-AC02-07CH11359 with the US Department of Energy.

\begin{appendix}

\section{$3 \to 2$ Scattering Rates}
\label{sec:app1}

In this appendix, we will derive a general form for $3 \to 2$ scattering rates, $\sigma v^2 (\, X_1 \, X_2 \, X_3 \to X_1^\prime \, X_2 ^\prime \, )$. Let $| i \rangle$ and $|f \rangle$ abbreviate the initial and final states, respectively. The relevant matrix element is related to the amplitude, $i \mathcal{M}$, by
\be
\langle f | i \rangle = (2 \pi)^4 ~ \delta^4 (k_\text{in} - k_\text{out}) ~ i \mathcal{M}
~,
\ee
where $k_\text{in,out}^\mu$ is the total incoming or outgoing 4-momenta. The probability, $P$, for this process to occur is then given by
\be
P = \frac{|\langle f | i \rangle|^2}{\langle f | f \rangle \langle i | i \rangle}
~.
\ee
Imagining that the scattering occurs in a spacetime box of spatial volume $V$ and time $T$, the numerator above can then be written as
\begin{align}
|\langle f | i \rangle|^2 &= \left[ (2 \pi)^4 ~ \delta^4 (k_\text{in} - k_\text{out}) \right]^2 ~ |\mathcal{M}|^2 
\nl
&= (2 \pi)^4 ~ \delta^4 (k_\text{in} - k_\text{out}) ~ (2 \pi)^4 ~ \delta^4 (0) ~  |\mathcal{M}|^2 
\nl
&= (2 \pi)^4 ~ \delta^4 (k_\text{in} - k_\text{out}) ~ V \, T ~  |\mathcal{M}|^2
~.
\end{align}
The single-particle states are normalized as
\be
\langle k | k \rangle = (2 \pi)^3 ~ 2 E_k ~ \delta^3(0) = 2 ~ E_k ~ V
~.
\ee
Therefore, $\dot{P} \equiv P / T$ can be expressed as
\be
\dot{P} = \frac{ (2 \pi)^4 ~ \delta^4 (k_\text{in} - k_\text{out}) ~ |\mathcal{M}|^2}{8 E_1 E_2 E_3 \times 4 E_1^\prime E_2^\prime \times  V^4 } 
~.
\ee
Summing over the outgoing momenta results in a factor of $V \times d^3 k_i^\prime / (2 \pi)^3$ for each outgoing particle. This gives
\be
\dot{P} = \frac{(2 \pi)^4 ~ \delta^4 (k_\text{in} - k_\text{out}) ~ |\mathcal{M}|^2}{8 E_1 E_2 E_3 \times V^2} ~~ \widetilde{dk_1^\prime} ~ \widetilde{dk_2^\prime}
~,
\ee
where $\widetilde{dk_i} \equiv d^3k_i / (2 \pi)^3 2 E_i$~. ``$\sigma$" is defined such that $\sigma \equiv \dot{P} / \text{flux}$~. Therefore,
\be
\sigma = \frac{\dot{P}}{(v_1 / V)(v_2 / V)}
~,
\ee
and hence
\be
\sigma v^2 = \dot{P} ~ V^2 =  \frac{(2 \pi)^4 ~ \delta^4 (k_\text{in} - k_\text{out}) ~ |\mathcal{M}|^2}{8 E_1 E_2 E_3} ~~ \widetilde{dk_1^\prime} ~ \widetilde{dk_2^\prime}
~.
\ee
In the non-relativistic limit, $E_i \approx m_i$, 
\be
\sigma v^2 = \frac{1}{8 m_1 m_2 m_3} \int d\text{LIPS}_2 ~ |\mathcal{M}|^2
~,
\ee
where $d\text{LIPS}_2 \equiv (2 \pi)^4 ~ \delta^4 (k_\text{in} - k_\text{out}) ~~ \widetilde{dk_1^\prime} ~ \widetilde{dk_2^\prime}$~. Also in the non-relativistic limit, the phase space integral is evaluated to be
\begin{align}
d\text{LIPS}_2 = \frac{d \cos{\theta}}{16 \pi ~ (m_1 + m_2 + m_3)^2} ~ \Big[ (m_1 + m_2 + m_3)^4 - 2 (m_1 + m_2 + m_3)^2 ~ (m_{1^\prime}^2 + m_{2^\prime}^2 ) +  (m_{1^\prime}^2 - m_{2^\prime}^2 )^2 \Big]^{1/2}
~. \nonumber \\
\end{align}
Therefore, we find
\be
\sigma v^2 = \frac{\Big[ (m_1 + m_2 + m_3)^4 - 2 (m_1 + m_2 + m_3)^2 ~ (m_{1^\prime}^2 + m_{2^\prime}^2 ) +  (m_{1^\prime}^2 - m_{2^\prime}^2 )^2 \Big]^{1/2}}{S \times 128 \pi ~ m_1 m_2 m_3 ~ (m_1 + m_2 + m_3)^2} \int_{-1}^1 d \cos{\theta} \, |\mathcal{M}|^2 ~, \nonumber \\
\ee
where $S$ is a symmetry factor for identical outgoing states. In the limit that all incoming and outgoing particles are mass degenerate, $m_{1,2,3} = m_{1^\prime, 2^\prime} = m$, this reduces to
\be
\frac{\sqrt{5}}{S \times 384 \pi ~ m^3} \int_{-1}^1 d \cos{\theta} ~ |\mathcal{M}|^2
~,
\ee
in agreement with that presented in Ref.~\cite{Choi:2015bya}.

\end{appendix}

\bibliography{hidden}

\end{document}